\documentclass[%
 aip,
 amsmath,amssymb,
preprint,%
]{revtex4-1}

\usepackage{graphicx}% Include figure files
\usepackage{dcolumn}% Align table columns on decimal point
\usepackage{bm}% bold math
\usepackage[utf8]{inputenc}
\usepackage[T1]{fontenc}
\usepackage{mathptmx}
\usepackage{makecell}

\begin{document}

% \preprint{AIP/123-QED}

\title[]{Friction Factor for Steady Periodically Developed Flow in Micro- and Mini-Channels with Arrays of Offset Strip Fins}
% \title[]{Steady Periodically Developed Flow and Heat Transfer in Micro- and Mini-Channels with Arrays of Offset Strip Fins: Part I - Friction Factor}
% \title[Macro-Scale Closure Correlations for Flow Through Arrays of Offset Strip Fins in Micro- and Mini-Channels]{Macro-Scale Closure Correlations for Flow Through Arrays of Offset Strip Fins in Micro- and Mini-Channels}
% \title[Low Reynolds Number Flow Regimes and the Use of Unit Cell Methods for Offset Strip Fin Compact Heat Exchangers]{Low Reynolds Number Flow Regimes and the Use of Unit Cell Methods for Offset Strip Fin Compact Heat Exchangers}
% Force line breaks with \\

\author{A. Vangeffelen}
\email{arthur.vangeffelen@kuleuven.be.}
\author{G. Buckinx}%
\affiliation{ 
Department of Mechanical Engineering, KU Leuven, Celestijnenlaan 300A, 3001 Leuven, Belgium%\\This line break forced with \textbackslash\textbackslash
}%
\affiliation{%
VITO, Boeretang 200, 2400 Mol, Belgium%\\This line break forced% with \\
}%
\affiliation{%
EnergyVille, Thor Park, 3600 Genk, Belgium%\\This line break forced% with \\
}%

\author{M. R. Vetrano}
\affiliation{ 
Department of Mechanical Engineering, KU Leuven, Celestijnenlaan 300A, 3001 Leuven, Belgium%\\This line break forced with \textbackslash\textbackslash
}%
\affiliation{%
EnergyVille, Thor Park, 3600 Genk, Belgium%\\This line break forced% with \\
}%

\author{M. Baelmans}
\affiliation{ 
Department of Mechanical Engineering, KU Leuven, Celestijnenlaan 300A, 3001 Leuven, Belgium%\\This line break forced with \textbackslash\textbackslash
}%
\affiliation{%
EnergyVille, Thor Park, 3600 Genk, Belgium%\\This line break forced% with \\
}%

\date{\today}

\begin{abstract}
In this work, the friction factor for steady periodically developed flow through micro- and mini-channels with periodic arrays of offset strip fins is analyzed. 
The friction factor is studied numerically on a unit cell of the array for Reynolds numbers ranging from 1 to 600, and fin height-to-length ratios below 1. 
It is shown that the friction factor correlations from the literature, which primarily focus on larger conventional offset strip fin geometries in the transitional flow regime, do not predict the correct trends for laminar flow in micro- and mini-channels. 
Therefore, a new friction factor correlation for micro- and mini-channels with offset strip fin arrays is constructed from an extensive set of numerical simulations through a least-squares fitting procedure. 
The suitability of this new correlation is further supported by means of the Bayesian approach for parameter estimation and model validation. 
The correlation predicts an inversely linear relationship between the friction factor and the Reynolds number, in accordance with our observation that a strong inertia regime prevails over nearly the entire range of investigated Reynolds numbers. 
Yet, through a more detailed analysis, also the presence of a weak inertia regime and a transitional regime is identified, and the transitions from the strong inertia regime are quantified by means of two critical Reynolds numbers. 
Finally, the new correlation also incorporates the asymptotic trends that are observed for each geometrical parameter of the offset strip fin array, and whose origins are discussed from a physical perspective. \\

\end{abstract}

\maketitle

\section{\label{sec:intro}Introduction}

Over the last two decades, considerable research has been devoted to flow through micro- and mini-channels with a periodic array of solid fin structures (Refs.~\onlinecite{kandlikar2005heat,khan2006role,izci2015effect,yang2017heatpin}). 
Typically, micro-channels are defined as channels with the smallest dimension between 10 $\mu$m and 200 $\mu$m, whereas for mini-channels, the smallest dimension ranges from 200 $\mu$m to 3 mm (Refs.~\onlinecite{kandlikar2005heat}). 
In particular, micro- and mini-channels with periodic arrays of offset strip fins have gained a lot of interest (Refs.~\onlinecite{bapat2006thermohydraulic,yang2007advanced, hong2009three,do2016experimental,nagasaki2003conceptual,yang2017heat,jiang2019thermal,yang2014design,pottler1999optimized}). 
This is primarily due to the compactness this type of fin surface enables for heat transfer devices, which offers great potential in the ongoing trend towards higher power densities in many applications (Refs.~\onlinecite{yang2007advanced,hong2009three}). 
For example, micro-channels with a periodic array of offset strip fins have been applied for the cooling of microelectronic devices, such as high-performance processors and vertically stacked integrated circuit chips (Refs.~\onlinecite{tuckerman1981high, bapat2006thermohydraulic, yang2007advanced, hong2009three}).
The reason is that they provide an effective means to cope with the increasing heat dissipation in these electronic devices as a result of their continuous miniaturization (Refs.~\onlinecite{bartolini2012thermal}). 
Additionally, mini-channels with offset strip fins are used in applications such as heat recuperators for compact gas turbines (Refs.~\onlinecite{do2016experimental,nagasaki2003conceptual}), cryogenic heat exchangers for refrigeration and liquefaction (Refs.~\onlinecite{yang2017heat,jiang2019thermal}), and solar air heaters (Refs.~\onlinecite{yang2014design,pottler1999optimized}). 
Here, the high thermal performance of offset strip fins allows for compact designs or small temperature differences between the two flows. \\

Due to the small dimensions, the flow inside micro- and mini-channels mostly remains in the laminar regime. It is characterized by a low to moderate Reynolds number, which typically lies between 10 and 500 (Refs.~\onlinecite{tuckerman1981high, bapat2006thermohydraulic, yang2007advanced, hong2009three,do2016experimental,nagasaki2003conceptual,yang2017heat,jiang2019thermal,yang2014design,pottler1999optimized,zargartalebi2020impact,gluzdov2021friction}). 
Inside arrays of offset strip fins, the flow exhibits a transition to vortex shedding in the wakes of the most downstream fins for Reynolds numbers between 350 and 750, depending on the geometrical parameters (Refs.~\onlinecite{joshi1987heat,mochizuki1988flow,dejong1997experimental}). % 350: mochizuki1988flow p. 1371, 750: joshi1987heat p. 79 (both D_hyd)
The onset point of this transition travels upstream in the channel as the Reynolds number is increased, until the oscillating flow breaks up and the flow becomes chaotic throughout the entire channel (Refs.~\onlinecite{mochizuki1988flow,dejong1997experimental}). 
Similar observations have been made for plate channels with pin-fin arrays, in which the transition occurs at Reynolds numbers between 250 and 850 (Refs.~\onlinecite{renfer2011experimental,xu2018experimental}). 
The flow regime inside micro- and mini-channels with offset strip fin arrays is therefore often steady. 
% Additionally, when arrays of offset strip fins are employed, the fin height is usually kept small relative to the fin length (Refs.~\onlinecite{tuckerman1981high, bapat2006thermohydraulic, yang2007advanced, hong2009three,do2016experimental,nagasaki2003conceptual,yang2017heat,jiang2019thermal,yang2014design,pottler1999optimized}). 

In the past, the flow through periodic arrays of offset strip fins has been studied both experimentally and numerically (Refs.~\onlinecite{joshi1987heat,manglik1995heat,dong2007air,kim2011correlations}). 
However, previous work mainly focuses on conventional channels with dimensions in the centimeter range, used for example in automotive radiators and air coolers.
As a consequence, most of the data in the literature only applies to the transitional and turbulent flow regime, but not to the steady flow regime at low to moderate Reynolds numbers, as encountered in micro- and mini-channels. 
Moreover, the available data mainly applies to offset strip fins with larger fin height-to-length ratios than those common in micro- and mini-channel applications (Refs.~\onlinecite{tuckerman1981high, bapat2006thermohydraulic, yang2007advanced, hong2009three,do2016experimental,nagasaki2003conceptual,yang2017heat,jiang2019thermal,yang2014design,pottler1999optimized}). 
Therefore, this work aims to study the pressure drop for flow through offset strip fin arrays in micro- and mini-channels. \\

\begin{figure}
\includegraphics[scale = 1.0]{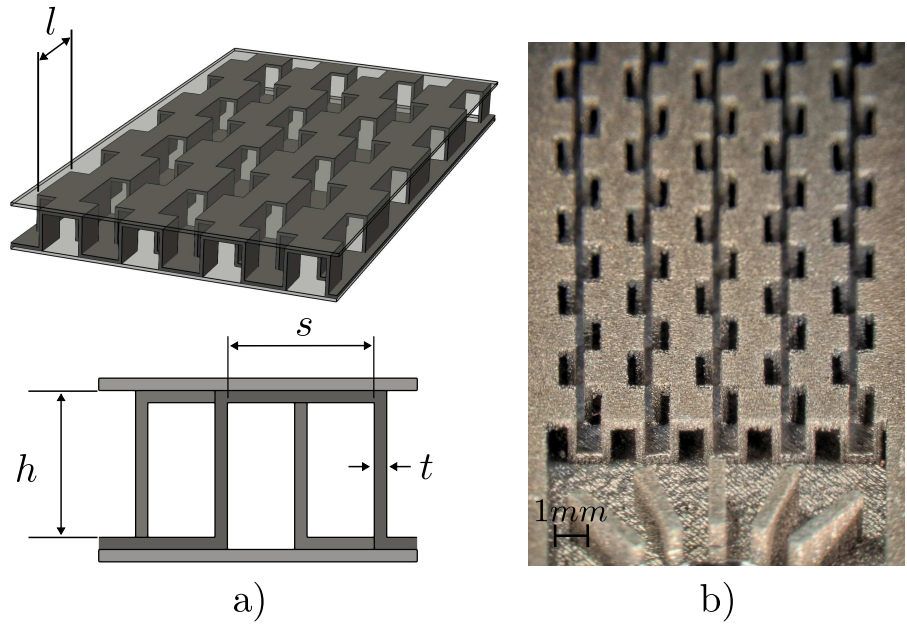}
\caption{\label{fig:osf} Geometrical parameters of a periodic array of offset strip fins (a),  example of a mini-channel with an array of periodic offset strip fins, fabricated in-house by the additive manufacturing technique of laser powder bed fusion (b)}
\end{figure}

The data from the literature consists to a large extent of pressure drop measurements over the array for different flow rates through the channel. 
These pressure drop measurements, often corrected for in-and outlet contractions, are frequently represented in the form of a friction factor, for which empirical correlations have been proposed. 
Tables \ref{tab:literature} and \ref{tab:literature2} give a chronological overview of the different friction factor correlations for offset strip fins reported in the literature. 
In the literature, the friction factor has been defined as the average Fanning friction factor $f \triangleq \frac{\mathrm{\Delta P} D_h}{2 L \rho_f U_{\text{ref}}^{2}}$, which is based on the pressure drop $\mathrm{\Delta P}$ measured over the flow length $L$. 
The overview further includes the adopted definition of the hydraulic diameter $D_h$ and the range of Reynolds numbers $Re_{D_h}\triangleq \rho_f U_{\text{ref}} D_h/\mu_f$ based on this hydraulic diameter, as well as the ratio of fin height to fin length $h/l$ considered in each study. 
The reference velocity $U_{\text{ref}}$ equals the average bulk velocity based on the minimal cross-section area $sh$ for all included correlations, with the exception of the correlation of Joshi and Webb, for which it is based on the surface area $(s-t)h$. 
Figure \ref{fig:osf} displays the geometrical parameters of an array of offset strip fins with their conventional symbols: the fin thickness $t$, the fin height $h$, the lateral fin pitch $s$, and the fin length $l$. Typically, the lateral offset $\sigma$ between successive fin rows is taken to be half of the lateral fin pitch: $\sigma = \left( s+t \right)/2$ (Refs.~\onlinecite{kays1984compact,manson1950correlations,wieting1975empirical,joshi1987heat,manglik1995heat,dong2007air,guo2008lubricant,kim2011correlations}). 
Figure \ref{fig:osf} further presents a real example of an offset strip fin mini-channel which has been produced in-house by the laser powder bed fusion manufacturing technique (Refs.~\onlinecite{jadhav2021laser}). 
Next, a short discussion on the studies from Tables \ref{tab:literature} and \ref{tab:literature2} is given. \\

%% accuracy f correlations
% Manson: Overestimation of f with factor 2
% Wieting: 85% of data within 15%
% Joshi: rms 13%, 82% of data within 15%
% Manglik: within 20% 
% Dong: 90% of data within 10%, mean 5.3%
% Guo: 85% of data within 15%, mean 5.73%
% Kim: R2 0.98, within 20% 

\begin{table*}[ht!]
\caption{\label{tab:literature}Chronological list of flow correlation studies for offset strip fin channels}
\begin{ruledtabular}
\begin{tabular}{llcc}
Researchers & Correlation & $Re_{D_h}$ & $h/l$ \\
\hline\hline
\makecell[l]{Manson \\ (1950)} & \makecell[l]{
    For $Re_{D_h} \leqslant 3500$ \\ 
    \quad $f = 
    \begin{cases}
        11.8 (l/D_h)^{-1} Re_{D_h}^{-0.67} & \text{for } l/D_h \leqslant 3.5 \\
        11.8 (3.5)^{-1}   Re_{D_h}^{-0.67} & \text{for } l/D_h > 3.5
    \end{cases} $\\
    For $Re_{D_h} > 3500$ \\ 
    \quad $f = 
    \begin{cases}
        0.38 (l/D_h)^{-1} Re_{D_h}^{-0.24} & \text{for } l/D_h \leqslant 3.5 \\
        0.38 (3.5)^{-1}   Re_{D_h}^{-0.24} & \text{for } l/D_h > 3.5
    \end{cases} $\\
    where $\quad D_h = 2sh/(s+h)$}
    & 370-9020 & 4.8-7.9\\
\hline
\makecell[l]{Wieting \\ (1975)} & \makecell[l]{
    For $Re_{D_h} \leqslant 1000$ \\ 
    \quad $f = 7.661 (l/D_h)^{-0.384} (s/h)^{-0.092} Re_{D_h}^{-0.712}$ \\
    For $Re_{D_h} > 2000$ \\ 
    \quad $f = 1.136 (l/D_h)^{-0.781} (t/D_h)^{0.534} Re_{D_h}^{-0.198}$ \\
    where $D_h = 2sh/(s+h)$}
    & 120-50000 & 0.23-5.1\\
\hline
\makecell[l]{Joshi \& Webb \\ (1987)} & \makecell[l]{
    For $Re_{D_h} < Re^{*}$ \\ 
    \quad $f = 8.12 (l/D_h)^{-0.41} (s/h)^{-0.02} Re_{D_h}^{-0.74}$ \\
    For $Re_{D_h} > Re^{*} + 1000$ \\ 
    \quad $f = 1.12 (l/D_h)^{-0.65} (t/D_h)^{0.17} Re_{D_h}^{-0.36}$ \\
    where \\
    \quad $Re^{*} = 257 (l/s)^{1.23} (t/l)^{0.58} D_h$ \\
    \quad \quad $\times \left[ t + 1.328 \left( Re_{D_h} / (l D_h) \right)^{-0.5}  \right]^{-1} $ \\
    % with $Re^{*} = \frac{2.57 (l/s)^{1.23} (t/l)^{0.58} D_h}{t + 1.328 \left( Re_{D_h} / (lD_h) \right)^{-0.5}} $ \\
    \quad $D_h = 2(s-t)h/((s+h)+ht/l))$}
    % \quad $U_{ref}$ is the average velocity based on the area $(s-t)h$
    & 120-50000 & 0.23-5.1\\
\hline
\makecell[l]{Manglik \& Bergles \\ (1995)} & \makecell[l]{
    $f = 9.6243 (s/h)^{-0.1856} (t/l)^{0.3053} (t/s)^{-0.2659} Re_{D_h}^{-0.7422}$ \\
    \quad $\times \left[ 1 + 7.669 \cdot 10^{-8} (s/h)^{0.920} (t/l)^{3.767} (t/s)^{0.236} Re_{D_h}^{4.429} \right]^{0.1} $ \\
    where $D_h = 4shl/(2(sl+hl+th)+ts)$}
    & 120-10000 & 0.23-5.1\\
\end{tabular}
\end{ruledtabular}
\end{table*}

\begin{table*}[ht!]
\caption{\label{tab:literature2}Chronological list of flow correlation studies for offset strip fin channels (continued)}
\begin{ruledtabular}
\begin{tabular}{llcc}
Researchers & Correlation & $Re_{D_h}$ & $h/l$\\
\hline\hline
\makecell[l]{Dong, Chen et al. \\ (2007)} & \makecell[l]{
    $f = 2.092 (s/h)^{-0.739} (t/l)^{-0.78} (t/s)^{0.972} Re_{D_h}^{-0.281} (L/l)^{-0.497}$ \\
    where \\
    \quad $L$ the flow length \\
    \quad $D_h = 2sh/(s+h)$}
    & 500-7500 & 0.91-2.3\\
\hline
\makecell[l]{Guo, Qin et al. \\ (2008)} & \makecell[l]{
    $f = 0.9415 (2 (s+t)/D_h)^{-1.052 \alpha + 6.161} (h/D_h)^{1.0540 \alpha + 2.057}$ \\
    \quad $\times (l/D_h)^{0.3388 \alpha + 0.4825 } Re_{D_h}^{-0.06420} F$ \\
    where \\
    \quad  $F = 0.05166 - 0.4983/\alpha + 1.342/\alpha^{2} - 1.002/\alpha^{3}$ \\
    \quad \quad $+ 0.1254/Re_{D_h}^{0.5166} $\\
    \quad $\alpha$ is the flow angle of attack \\
    \quad $D_h = 4V_{free}/A_{surface}$ \\
    \quad $V_{free}$ is the free flow volume \\
    \quad $A_{surface}$ is the wetted surface area}
    & 20-400 & 1.9-2.9\\ % 2sh/(s+h)
\hline
\makecell[l]{Kim, Lee et al. \\ (2011)} & \makecell[l]{
    For $b < 0.2$ \\
    \quad $f = exp(7.91) (s/h)^{-0.159} (t/l)^{0.358} (t/s)^{-0.033} Re_{D_h}^{0.126 ln (Re_{D_h}) - 2.3}$ \\
    For $0.2 \leqslant b < 0.25$ \\
    \quad $f = exp(9.36) (s/h)^{-0.0025} (t/l)^{-0.0373} (t/s)^{1.85} Re_{D_h}^{0.142 ln (Re_{D_h}) - 2.39}$ \\
    For $0.25 \leqslant b < 0.3$ \\
    \quad $f = exp(5.58) (s/h)^{-0.36} (t/l)^{0.552} (t/s)^{-0.521} Re_{D_h}^{0.111 ln (Re_{D_h}) - 1.87}$ \\
    For $0.3 \leqslant b < 0.35$ \\
    \quad $f = exp(4.84) (s/h)^{-0.48} (t/l)^{0.347} (t/s)^{0.511} Re_{D_h}^{0.089 ln (Re_{D_h}) - 1.49}$ \\
    where \\
    \quad $b = ((2s + 2t)(h + t) - 2sh)/((2s + 2t)(h + t))$\\
    \quad $D_h = 4shl/(2(sl+hl+th)+ts)$}
    & 100-6000 & 0.046-10\\
\end{tabular}
\end{ruledtabular}
\end{table*}

\newpage
\clearpage

The first friction factor correlation was constructed by Manson (Refs.~\onlinecite{manson1950correlations}). 
It was based on experimental pressure drop measurements for nine different types of heat transfer surfaces, all consisting of periodic fin arrays, of which only three were offset strip fin geometries (Refs.~\onlinecite{manson1950correlations}). 
As it can be seen from the range of Reynolds numbers and fin height-to-length ratios in Table \ref{tab:literature}, the experimental data in Manson's study is limited to flow regimes beyond the transition towards vortex shedding and to larger conventional fin geometries. 

Following Manson's work, numerous other correlations have been constructed, for which the experimental pressure drop measurements of Kays and London served as the main source of data (Refs.~\onlinecite{kays1984compact}). 
However, the database of Kays and London has limited applicability to micro- and mini-channels with offset strip fins, since only 13 of the 21 geometries in the database comply with $h/l \leq 1$. Moreover, less than 20\% of the measurement points in the database relates to the steady laminar flow regime. 

First, Wieting (Refs.~\onlinecite{wieting1975empirical}) proposed a friction factor correlation for offset strip fins using the database of Kays and London, albeit without giving a consistent definition of the hydraulic diameter (Refs.~\onlinecite{manglik1995heat}). 
A second correlation based on the data of Kays and London was constructed by Joshi and Webb (Refs.~\onlinecite{joshi1987heat}), who also included data for two geometries of Walters (Refs.~\onlinecite{walters1969hypersonic}), another geometry of London and Shah (Refs.~\onlinecite{shah1967offset}), as well as eight of their own geometries.
Of these eight additional geometries, only one has a fin height-to-length ratio smaller than one. 
In total, less than 20\% of Joshi and Webb's data points correspond to steady laminar flow conditions with $h/l \leq 1$. 
It must be remarked that Joshi and Webb also worked out a semi-analytical expression for the Reynolds number $Re^{*}$ at which transition from a laminar to a turbulent flow regime occurs (see Table \ref{tab:literature2}). 

A refit of the combined friction factor data of Kays and London, Walters, and London and Shah (Refs.~\onlinecite{kays1984compact,walters1969hypersonic,shah1967offset}) was performed by Manglik and Bergles (Refs.~\onlinecite{manglik1995heat}).
Manglik and Bergles' refitted correlation is perhaps the most widely used correlation for offset strip fins in conventional channels. 
One reason is that it is accurate to within a relative error of 20\% with respect to the data considered in their work.
Secondly, it also covers a wide range of Reynolds numbers with a single formula, without any partitioning around the transition point towards the turbulent flow regime. 
Considering that the underlying data comes from the previous studies, the refit of Manglik and Bergles has again limited applicability to micro- and mini-channels with offset strip fins.

For offset strip fin arrays in channels of a rather short length, the pressure drop over the entrance region has been included in the friction factor correlation of Dong, Chen et al. (Refs.~\onlinecite{dong2007air}) through the ratio of the flow length $L$ and the fin length $l$, as shown in Table \ref{tab:literature2}. 
To this end, pressure drop measurements were experimentally conducted for 16 different offset strip fin arrays containing 5 to 14 fins in the flow direction. 
Yet, since the experimental data was gathered for the transitional flow regime in fins with a relatively large height, as it can be seen in Table \ref{tab:literature2}, the validity of the latter study for the micro- and mini-channels considered in this work is still a question. 

In contrast to the previous studies, a significant number of data points for the laminar flow regime have been incorporated in the friction factor correlation of Guo, Qin et al. (Refs.~\onlinecite{guo2008lubricant}). 
The experimental study of Guo, Qin et al. introduced offset strip fin arrays having an inclined direction with respect to the flow, and was aimed at evaluating their performance for a novel design of, for instance, oil coolers (Refs.~\onlinecite{muzychka2001modeling,zhang2015numerical}).
To this end, the authors investigated the dependence of the pressure drop on the direction of the flow, i.e. the angle of attack, with respect to the fin orientation. 
Therefore, their data was gathered from experiments for 16 combinations of offset strip fin geometries and flow angles ranging from 0$^{\circ}$ to 90$^{\circ}$. 
Unfortunately, in their work, neither an explicit definition of the hydraulic diameter nor a value of the fin thickness are given, so that the interpretation of their friction factor correlation remains ambiguous. 
Furthermore, only larger fin height-to-length ratios ($h/l > 1$) than common for micro- and mini-channels were included in their data set, as it is clear from Table \ref{tab:literature2}. 

Finally, the only numerical study on the friction factor for arrays of offset strip fins has been conducted by Kim, Lee et al. (Refs.~\onlinecite{kim2011correlations}). 
In this study, the friction factor was computed from flow simulations on a three-dimensional domain consisting of an entire fin row spanning 68 fin lengths in the streamwise direction, under the assumption of lateral flow periodicity.
However, as the numerical data from this study has not been published, it is unclear how many friction factor data points were included to cover the laminar flow regime.
Furthermore, it seems that this study was rather focused on the turbulent flow regime, since the flow simulations were performed using the shear stress transport (SST) k-$\omega$ turbulence model to solve the flow equations for an incompressible Newtonian fluid. 
The empirical correlation constructed by Kim, Lee et al. is based on the numerical data for 39 different geometries, of which 30 have a fin height-to-length ratio $h/l$ smaller than one. 
In particular, the correlation covers a large range of values for the so-called blockage ratio $b$, which is defined in Table \ref{tab:literature2} as the cross-section area blocked by the fins over the total cross-section area in the channel. \\

\newpage
\clearpage

\begin{figure}
\includegraphics[scale = 1.0]{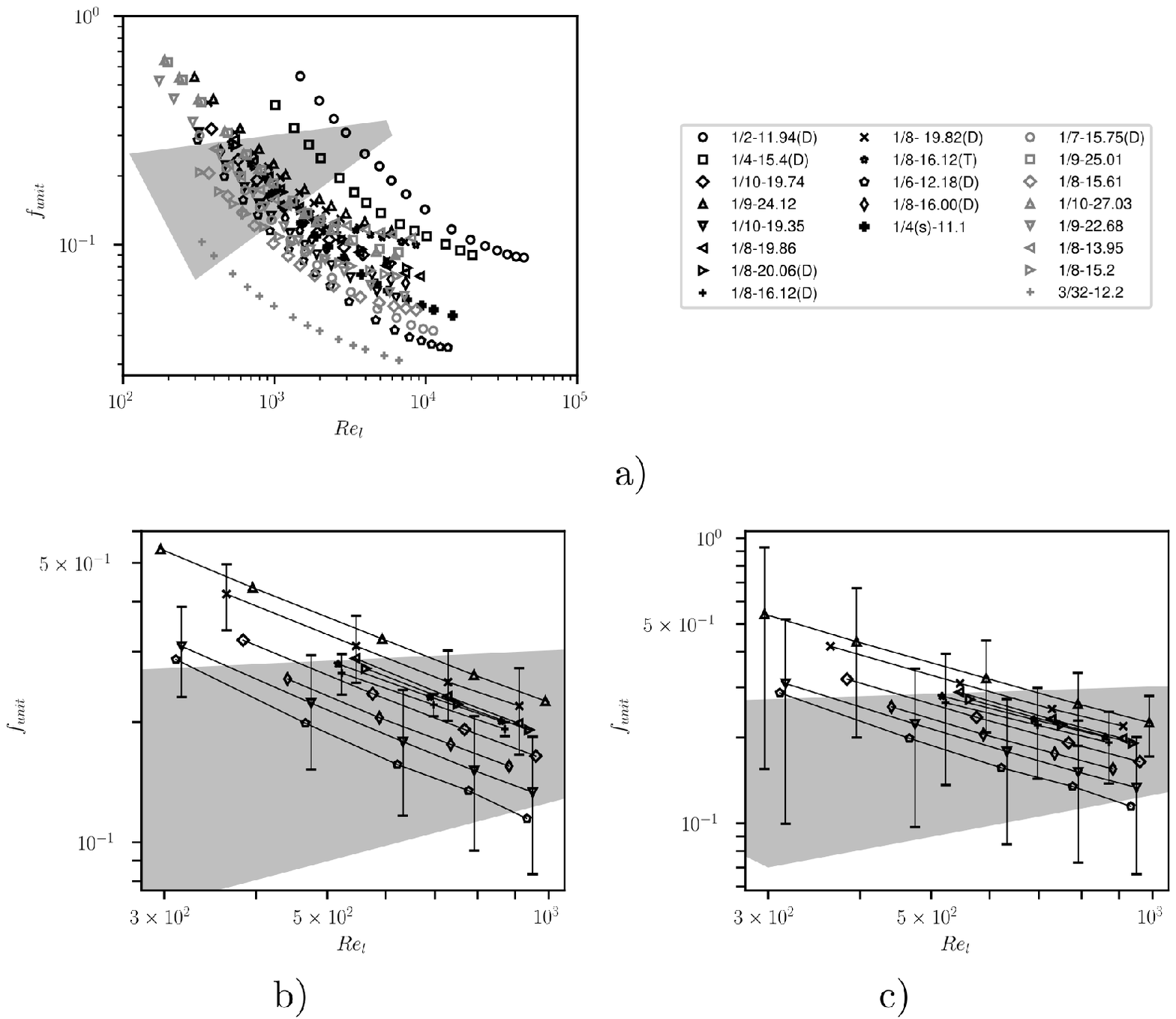}
\caption{\label{fig:literature_comparison} Experimental friction factor data from Kays and London (Refs.~\onlinecite{kays1984compact}) (a), discrepancies between the data of Kays and London and the friction factor correlation of Manglik and Bergles (Refs.~\onlinecite{manglik1995heat}) (b), discrepancies between the data of Kays and London and the friction factor correlation of Kim, Lee et al. (Refs.~\onlinecite{kim2011correlations}) (c); The black markers correspond to the geometries with $h/l < 1$, and the grey markers to those with $h/l > 1$.
The grey zone indicates the region of transitional flow.}
\end{figure}

From the preceding literature review, we conclude that up to the present, the friction factor for the steady laminar flow regime in offset strip fin arrays as encountered in micro- and mini-channels is not well studied. 
After all, in total less than 50 pressure drop measurement points are available for the laminar regime in the entire literature. 
To support this conclusion graphically,
almost all the experimental friction factor measurements in the literature, coming from the database of Kays and London (Refs.~\onlinecite{kays1984compact}), are displayed in Figure \ref{fig:literature_comparison}(a).
In this figure, the scarce measurement points for the laminar regime are located above the grey zone, which indicates the region of transitional flow. % corresponding to points in the transitional flow regime
This region extends more or less over the Reynolds numbers $Re_{D_h}$ between 350 and 750 (Refs.~\onlinecite{joshi1987heat,mochizuki1988flow,dejong1997experimental}). 
The limited measurement points for geometries with a small ratio of fin height to fin length ($h/l < 1$) are indicated by the black markers, while those with $h/l > 1$ are indicated by the grey markers.
In Figures \ref{fig:literature_comparison}(b) and \ref{fig:literature_comparison}(c) also the discrepancies between the data from Kays and London and the most commonly used correlations from the literature are shown by means of an error bar, at least
over the Reynolds number range relevant for micro- and mini-channels  (Refs.~\onlinecite{kays1984compact}). 
These discrepancies are on an average as large as 20\%, but become as large as 80\% for a Reynolds number $Re_{l}$ around 300. 

Therefore, in this work, the friction factor for the steady laminar flow regime inside offset strip fin arrays is re-investigated numerically and correlated with respect to the geometrical fin parameters for micro- and mini-channel applications ($h/l < 1$). 
In these applications, the flow is expected to become periodically developed after a short distance from the inlet (Refs.~\onlinecite{lee2005investigation}).
For that reason, the aim of this work is to quantify the friction factor for offset strip fins in the periodically developed flow regime. 
Since periodically developed flow can be simulated on a single unit cell of the fin array (Refs.~\onlinecite{patankar1977fully}), less computational resources are required than one would need for an analysis based on direct numerical simulation (DNS) of the entire flow in the channel.

The remainder of this work is organized as follows. 
First, in Section \ref{sec:period}, the unit cell geometry, the periodically developed flow equations and the numerical procedure are discussed. 
In Section \ref{sec:aligned}, the friction factor is analyzed in the periodically developed regime for a large set of geometrical parameters and Reynolds numbers ranging from 1 to 600. 
Based on the dependence of the friction factor on the Reynolds number, three different flow regimes are identified in Section \ref{sec:critical}: the weak inertial flow regime, the strong inertial flow regime and the transitional flow regime. 
To mark the transition between these regimes, two critical Reynolds numbers are determined. 
A final friction factor correlation for periodically developed flow through offset strip fin arrays in micro- and mini-channels is presented in Section \ref{sec:correlation}. 
Also its accuracy is evaluated with respect to our own data, as well as the experimental data from the literature. \\

\newpage
\clearpage

\section{\label{sec:period}Unit cell geometry and periodically developed flow equations}

\subsection{Geometry}

\begin{figure}[ht]
\includegraphics[scale = 1.0]{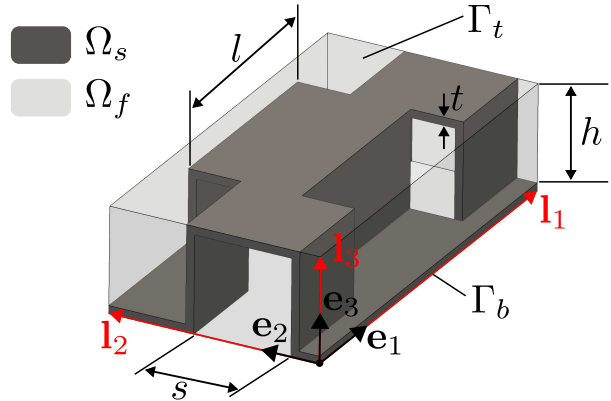}
\caption{\label{fig:UnitCellDomains} Unit cell domain in an offset strip fin array}
\end{figure}

The three-dimensional unit cell for the computation of the periodically developed flow through an offset strip fin array is depicted in Figure \ref{fig:UnitCellDomains}. 
The unit cell domain $\Omega_{unit} \subset \mathbb{R}^{3}$ exists of a fluid subdomain $\Omega_{f}$, a solid subdomain $\Omega_{s}$ and their fluid-solid interface $\Gamma_{fs}$. 
The top plane  $\Gamma_{t}$ and bottom plane $\Gamma_{b}$, which are both part of the exterior boundary of the unit cell domain $\Gamma = \partial \Omega_{unit}$, correspond to the solid walls at the top and bottom of the channel. 
With respect to the normalized Cartesian vector basis $\{ \textbf{e}_{j} \}_{j=1,2,3}$, the unit cell domain is spanned by three lattice vectors: $\textbf{l}_{1} = l_{1} \textbf{e}_{1} = 2l \textbf{e}_{1}$, $\textbf{l}_{2} = l_{2} \textbf{e}_{2} = 2(s+t) \textbf{e}_{2}$ and $\textbf{l}_{3} = l_{3} \textbf{e}_{3} = (h+t) \textbf{e}_{3}$. \\

Using the fin length $l$ as the reference length, the following independent non-dimensional groups are selected to uniquely define the unit cell geometry: $h/l$, $s/l$ and $t/l$. 
These non-dimensional parameters determine the unit cell's porosity as follows: 

\begin{equation}
    \epsilon = \frac{(h/l)(s/l)}{[(h/l)+(t/l)][(s/l)+(t/l)]}. 
\end{equation}

%\newpage
%\clearpage

\subsection{Periodically developed flow equations}

For laminar periodically developed flow, the pressure field is composed of a component varying linearly with a constant gradient $\mathrm{\nabla{P}}$ and a spatially periodic component $p^{*}$, which drive the periodic velocity field $\boldsymbol{u}$ (Refs.~\onlinecite{patankar1977fully,buckinx2015multi}). 
Assuming constant material properties, the periodically developed flow field $\boldsymbol{u}$ is the steady solution governed by the following time-dependent Navier-Stokes equations: 
\begin{equation}
\begin{aligned}
  \nabla \cdot \boldsymbol{u} &= 0 & \text{in } \Omega_{f},  \\
  \rho_{f} \frac{\partial \boldsymbol{u}}{\partial t} + \rho_{f} \nabla \cdot \left( \boldsymbol{u} \boldsymbol{u} \right) &= - \mathrm{\nabla{p}^{*}} - \mathrm{\nabla{P}} + \mu_{f} \nabla^2 \boldsymbol{u} & \text{in } \Omega_{f},
\end{aligned}
\label{eq:NavierStokes}
\end{equation}
where
\begin{equation}
\begin{aligned}
  \boldsymbol{u} \left( \textbf{x} + \textbf{l}_{j} \text{,} t \right) &= \boldsymbol{u} \left( \textbf{x} \text{,} t \right) & &\text{in } \Omega_{f}, \\
  p^{*} \left( \textbf{x} + \textbf{l}_{j} \text{,} t \right) &= p^{*} \left( \textbf{x} \text{,} t \right) & &\text{in } \Omega_{f}, \\
  \boldsymbol{u} \left( \textbf{x} \text{,} t \right) &= 0 & &\text{in } \Gamma_{fs} \cup \Gamma_{t} \cup \Gamma_{b}, \\
  \boldsymbol{u} \left( \textbf{x} \text{,} 0 \right) &= 0 & &\text{in } \Omega_{f}, \\
  p \left( \textbf{x} \text{,} 0 \right) &= 0 & &\text{in } \Omega_{f}, 
\end{aligned}
\label{eq:BoundaryConditions}
\end{equation}
for $j=\{ 1,2 \}$. 
Here, the no-slip condition is imposed at the fluid-solid interface $\Gamma_{fs}$, as well as at the top and bottom boundary $\Gamma_{t}$ and $\Gamma_{b}$, respectively. 
The flow further exhibits periodicity with respect to the lattice vectors $\textbf{e}_{1}$ and $\textbf{e}_{2}$. 
Given the periodically developed flow equations (\ref{eq:NavierStokes})-(\ref{eq:BoundaryConditions}), a developed pressure gradient $\mathrm{\nabla{P}}$ can be obtained for an imposed volume-averaged velocity vector $\langle \boldsymbol{u} \rangle$, which is defined as 
\begin{align}
  \langle \boldsymbol{u} \rangle  &\triangleq \frac{1}{V_{unit}} \int_{\textbf{r} \in \Omega_{unit} \left( \textbf{x} \right)} \boldsymbol{u} \left( \textbf{r} \text{,} t \right) \,d\Omega \left( \textbf{r} \right), \\
  V_{unit} & = \textbf{l}_{1} \cdot \left( \textbf{l}_{2} \times \textbf{l}_{3} \right). 
\end{align}

This relationship between $\mathrm{\nabla{P}}$ and $\langle \boldsymbol{u} \rangle$ will be represented in non-dimensional form as a correlation between the friction factor, defined as 
\begin{equation}
    f_{unit} \triangleq \frac{ \| \mathrm{\nabla{P}} \| l }{2 \rho_{f} \|\langle \boldsymbol{u} \rangle\|^2} \,, 
\label{eq:frictionfactor}
\end{equation}
and the Reynolds number, defined as
\begin{equation}
    Re_l \triangleq \frac{\rho_{f} \|\langle \boldsymbol{u} \rangle\| l}{\mu_{f}}, 
\label{eq:reynoldsnumber}
\end{equation}
with $\| \, \|$ denoting the Euclidean vector norm. 
Note that the fin length $l$ has been selected as the reference length for both the friction factor and Reynolds number, as opposed to the commonly used hydraulic diameter $D_{h}$. 
This choice of reference length is motivated by the fact that $\Vert \nabla P \Vert l$ corresponds to the constant pressure drop over each fin unit in the main flow direction, and the argument that a Reynolds number based on the same, single reference length is easier to interpret than a Reynolds number based on two geometrical parameters like $s$ and $h$.
Besides, the actual hydraulic diameter for a channel containing a fin array is not a constant in the strict sense, since the cross-sectional area for the flow varies along the main flow direction. 

When the periodically developed flow regime prevails over the largest part of the channel, the friction factor in equation \ref{eq:frictionfactor} will be a good approximation for the total pressure drop over the entire channel. 
Furthermore, when the influence of the channel's side-wall on the flow can be neglected, $\Vert \langle \boldsymbol{u} \rangle \Vert$ will equal the bulk-average velocity through the channel, and hence directly determine the mass flow rate through the channel. 

\subsection{Numerical procedure}

For solving the flow equations (\ref{eq:NavierStokes})-(\ref{eq:BoundaryConditions}), the software package FEniCSLab has been employed, which was developed by G. Buckinx within the finite-element framework FEniCS (Refs.~\onlinecite{AlnaesBlechta2015a}). 
The package FEniCSLab contains an object-oriented re-implementation of the parallel fractional-step solver of \textit{Oasis} developed by Mortensen and Valen-Sendstad (Refs.~\onlinecite{mortensen2015oasis}) for the unsteady incompressible Navier-Stokes equations but has been modified to allow for variable time stepping and to solve for an imposed value of $\langle \boldsymbol{u} \rangle$. 

A structured mesh has been used for the spatial discretization of the unit cell domain, with a number of grid cells equal to 48 along $\textbf{e}_{1}$, 26 to 108 along $\textbf{e}_{2}$, and 13 to 104 along $\textbf{e}_{3}$. 
This resulted in a total number of mesh cells between 127,088 and 4,066,816, depending on the geometrical parameters of the unit cell domain. 
The discretized velocity and pressure fields have been represented by continuous Galerkin tetrahedral elements of the second and first order, respectively. 
For the temporal discretization, a central difference scheme has been employed, while an implicit Crank-Nicolson scheme of the second order is used for the viscous term.
A combination of an explicit Adam-Bashforth scheme and an implicit Crank-Nicolson scheme, both of the second order, has been selected for the discretization of the advective term, analogous to the flow solver of \textit{Oasis} (Refs.~\onlinecite{mortensen2015oasis}). 

The numerical discretization scheme has been validated through a mesh-independence study at the two largest Reynolds numbers, for the four lowest values of the unit cell’s porosity. 
Stability for the time-stepping scheme was ensured by limiting the local Courant–Friedrichs–Lewy number in each mesh cell to a value below 0.9. 
As time-convergence criterion, the requirement that the $L^{2}$-norm of each of the dimensionless velocity components $u_i/\Vert \langle \boldsymbol{u} \rangle \Vert$ is smaller than $10^{-6}$ was chosen, next to the requirement that the relative change in friction factor is smaller than $10^{-6}$.
By means of the Richardson extrapolation method and the grid convergence index (Refs.~\onlinecite{richardson1911ix,roache1994perspective}), this mesh-refinement study has indicated that the relative discretization error on the computed friction factor does not exceed 1\%. 

\begin{figure}
\includegraphics[scale = 1.0]{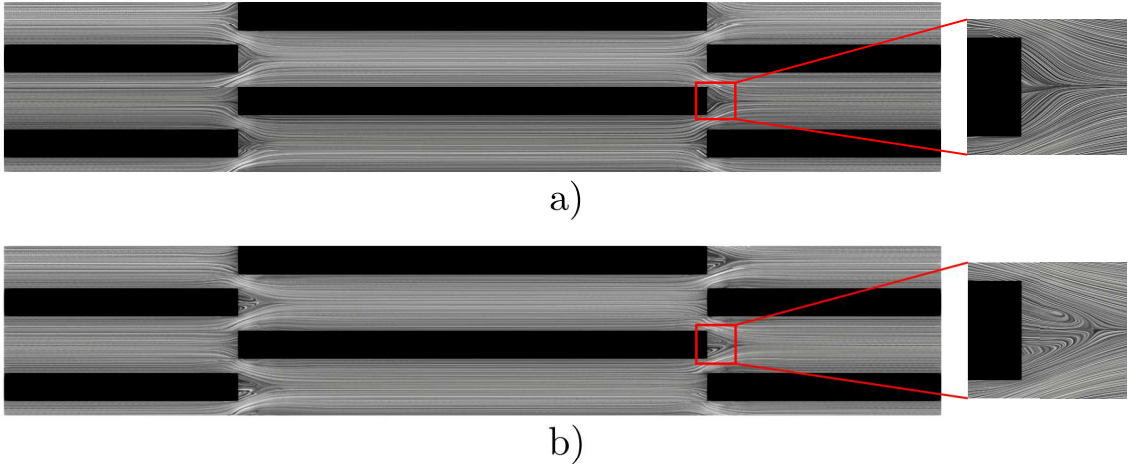}
\caption{\label{fig:streamlinesRE100_400T6H48S12} Example of flow pattern visualization by Line Integral Convolution (LIC) in a cross-section mid-plane spanned by $\textbf{e}_{1}$ and $\textbf{e}_{2}$ for $Re_{l}=100$ (a) and $Re_{l}=400$ (b) when $t/l=0.06$, $h/l=0.48$, $s/l=0.12$}
\end{figure}

As an example of the flow field inside a unit cell, the detailed flow patterns observed in the mid-plane of the unit cell spanned by $\textbf{e}_{1}$ and $\textbf{e}_{2}$ at a Reynolds number of 100 and 400 are shown in Figure \ref{fig:streamlinesRE100_400T6H48S12}. 
The growth of a wake with symmetrical recirculation zones at the fin's trailing edge can be seen for an increasing Reynolds number $Re_{l}$. 

% \newpage
% \clearpage

\section{\label{sec:aligned} Friction factor for periodically developed flow}

In order to investigate the relationship between $f_{unit}$ and $Re_{l}$, we consider the case where the volume-averaged velocity vector is parallel to the offset strip fins and thus aligned with the unit vector $\textbf{e}_{1}$. 
In that case, also the pressure gradient $\mathrm{\nabla{P}}$ and $\textbf{e}_{1}$ are aligned. 
For our study, we have selected the values of the geometrical parameters and the Reynolds number in accordance with the applications of offset strip fin arrays in micro- and mini-channels (Refs.~\onlinecite{bapat2006thermohydraulic,yang2007advanced, hong2009three,do2016experimental,nagasaki2003conceptual,yang2017heat,jiang2019thermal,yang2014design,pottler1999optimized}), as enlisted in Table \ref{tab:UCgeom}. 
These values of the geometrical parameters result in a porosity $\epsilon$ ranging from 0.44 to 0.97. 
The resulting total of 2765 data points for the friction factor $f_{unit}$, obtained by solving the periodically developed flow equations for 275 different geometries, are tabulated in Appendix \ref{sec:alignedflowdata}. 
% The resulting total of 2765 data points for the friction factor $f_{unit}$, obtained by solving the periodically developed flow equations for 275 different geometries, are available in (Refs.~\onlinecite{vangeffelen2021friction}). 
In the following subsections, the influence of the Reynolds number on the friction factor, as well as that of each geometrical parameter, are discussed. \\

\begin{table}[ht]
\caption{\label{tab:UCgeom} Adopted values for the Reynolds number and the geometrical parameters for our study of the friction factor for periodically developed flow.}
\begin{tabular}{l|l}
\hline\hline
$Re_{l}$ \quad & \quad 1, 10, 15, 25, 35, 50, 75, 100, 150, 200, 300, 400, 600\\
$h/l$    \quad & \quad 0.12, 0.16, 0.20, 0.24, 0.28, 0.32, 0.40, 0.48, 0.56, 0.68, 1.00\\
$s/l$    \quad & \quad 0.12, 0.16, 0.20, 0.24, 0.28, 0.32, 0.40, 0.48\\
$t/l$    \quad & \quad 0.01, 0.02, 0.04, 0.06\\
\hline\hline
\end{tabular}
\end{table}

\subsection{\label{sec:influence re}The influence of the Reynolds number $Re_l$ on the friction factor}

\begin{figure}[ht]
\centering
\begin{minipage}{.475\textwidth}
% \centering
\raggedleft
\includegraphics[scale = 1.00]{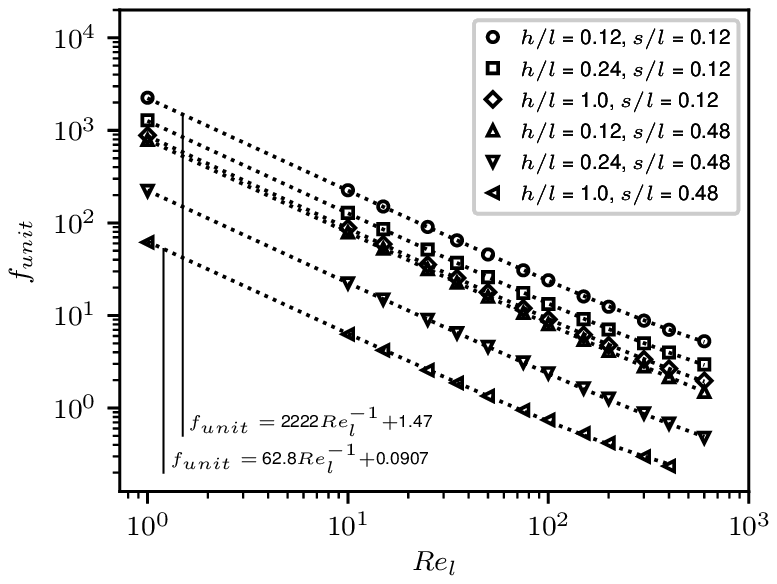}
\caption{\label{fig:REdataT4} Influence of the Reynolds number on the friction factor for steady periodically developed flow, when $t/l=0.04$ \newline\newline}
\end{minipage}
\hfill
\begin{minipage}{.475\textwidth}
% \centering
\raggedright
% \raggedleft
\includegraphics[scale = 1.00]{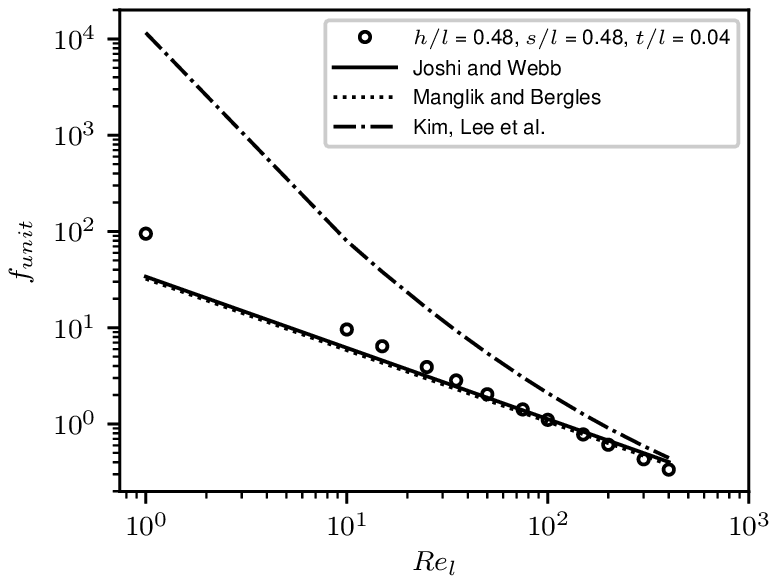}
\caption{\label{fig:litcompT4H48S48_nofit} A comparison between the friction factor correlations from the literature with the data from this work for steady periodically developed flow at low Reynolds numbers, when $t/l=0.04$, $h/l=0.48$, $s/l=0.48$}
\end{minipage}
\end{figure}

Figure \ref{fig:REdataT4} shows the dependence of the friction factor on the Reynolds number for various combinations of geometrical parameters. 
It can be observed that this dependence is well captured by an inversely linear relationship of the form
\begin{equation}
f_{unit} \simeq A Re_l^{-1} + B,
\label{eqn:ergun}
\end{equation}
where the parameters $A$ and $B$ and are only a function of the unit cell geometry. 
The exponent of the Reynolds number has been confirmed to effectively equal -1 with a standard deviation of 0.02 by means of a log-linear regression analysis of all the friction factor data from this work. 
For each unit cell geometry from table \ref{tab:UCgeom}, it was found that the correlation form (\ref{eqn:ergun}) captures the variation of the friction factor with the Reynolds number within an average and maximum relative error of 0.8\% and 8\%, respectively. 

In the correlation form (\ref{eqn:ergun}), the term $A Re_l^{-1}$ may be interpreted as originating from Darcy's law, which predicts a linear relationship between the pressure gradient and the volume-averaged velocity: $\Vert \nabla \mathrm{P} \Vert \sim \mu_f  A \Vert \langle \boldsymbol{u} \rangle \Vert$.
On the other hand, the constant $B$ in (\ref{eqn:ergun}) may be regarded as originating from Forchheimer's correction $\rho_f B \Vert \langle \boldsymbol{u} \rangle \Vert^2$ to Darcy's law, which is quadratic in the volume-averaged velocity
(Refs.~\onlinecite{forchheimer1901wasserbewegung,ergun1952fluid,ghaddar1995permeability,markicevic2002apparent,matsumura2014numerical}).
According to this interpretation, the correlation form (\ref{eqn:ergun}) reflects that periodically developed flow through an array of offset strip fins exhibits a regime of strong inertia, which is similar to the one encountered in steady laminar flows through porous media at low to moderate Reynolds numbers (Refs.~\onlinecite{koch1997moderate,lasseux2011stationary}).
In this regime of strong inertia within offset strip fins, the relationship between the pressure gradient and the volume-averaged velocity deviates only slightly from the linear relationship predicted by Darcy's law since the Forchheimer correction term remains relatively small, as illustrated in Figure \ref{fig:REdataT4}. 
Our observation that the Forchheimer coefficient $B$ has a limited contribution to the friction factor over the entire range of Reynolds numbers investigated here, is in agreement with the pressure drop measurements reported for other micro- and mini-channels applications (Refs.~\onlinecite{tuckerman1981high, bapat2006thermohydraulic, yang2007advanced, hong2009three,do2016experimental,nagasaki2003conceptual,yang2017heat,jiang2019thermal,yang2014design,pottler1999optimized}). 

Although the assumption in (\ref{eqn:ergun}) that the Forchheimer correction term scales quadratically with the volume-averaged velocity, leads to accurate friction factor correlations for our data (see Section \ref{sec:correlation}), and is supported by similar findings for flows through porous media and micro- and mini-channels, a more detailed analysis actually reveals a more complex dependence on the volume-averaged velocity. 
This more detailed analysis is discussed in Section \ref{sec:critical}.
Here, we first compare the accuracy of the  correlation form (\ref{eqn:ergun}) for our data with respect to the correlations proposed in the literature. 

In Figure \ref{fig:litcompT4H48S48_nofit}, the friction factor correlations from the literature (see Tables \ref{tab:literature} and \ref{tab:literature2}) are compared with the friction factor data from this work for a single geometry. 
Despite the fact that only a single geometry has been selected for comparison, this figure is also representative of the accuracy with which the correlations from the literature capture our data for other geometries.
We clarify that  the correlations from the literature, which are based on $f$ and $Re$, have been converted to our definitions of $f_{unit}$ and $Re_l$, using the definitions for $f$ and $Re$ specified in Section \ref{sec:intro}. 

As shown in Figure \ref{fig:litcompT4H48S48_nofit}, the friction factor correlations of Wieting, Joshi and Webb, and Manglik and Bergles (Refs.~\onlinecite{wieting1975empirical,joshi1987heat,manglik1995heat}) display a dependence on the Reynolds number of the form $f \sim Re^{-0.7}$, in contrast to the inversely linear relationship (\ref{eqn:ergun}) that we observed. 
Likely, the reason for this different exponent is that the largest part of the friction factor data used by the former authors covers the transitional and turbulent regime. 
It is known that towards the turbulent regime, the friction losses scale with $\Vert \langle \boldsymbol{u} \rangle \Vert^e$ with $e > 1$, so that $f \sim Re^{k}$ with $k > -1$ (Refs.~\onlinecite{ergun1952fluid}). 
As a result, the friction factor correlations of the former authors underestimate our friction factor data in the laminar regime with an average relative error of 20\% and a maximum relative error of 80\% for a Reynolds number near 1. 
An even larger underestimation of our data is apparent with respect to the study of Dong, Chen et al. (Refs.~\onlinecite{dong2007air}), who observed that $f \sim Re^{-0.281}$.
This can be explained by the fact that the validity range of their correlation was limited to a lowest Reynolds number equal to 500 (see Table \ref{tab:literature2}). 
In contrast, in the numerical study of Kim, Lee et al. (Refs.~\onlinecite{kim2011correlations}) also lower values of the Reynolds number were included in the data set. 
However, according to Kim, Lee et al., the friction factor scales approximately as $f \sim Re^{0.1 ln (Re) - 2}$. 
It can be seen from Figure \ref{fig:litcompT4H48S48_nofit} that the latter empirical correlation results in a significant overestimation of our friction factor data at Reynolds numbers $Re_l$ below 300, with a difference going up to a factor of one hundred as the Reynolds number decreases. 

The previous discussion thus shows that the correlations from the literature do not predict the appropriate scaling of the friction factor with the Reynolds number for steady laminar periodically developed flow in micro- and mini-channels. \\

% \newpage
% \clearpage

\subsection{\label{sec:influence h/l}The influence of the fin height-to-length ratio $h/l$ on the friction factor}

\begin{figure}[ht]
\centering
\begin{minipage}{.475\textwidth}
% \centering
\raggedleft
\includegraphics[scale = 1.00]{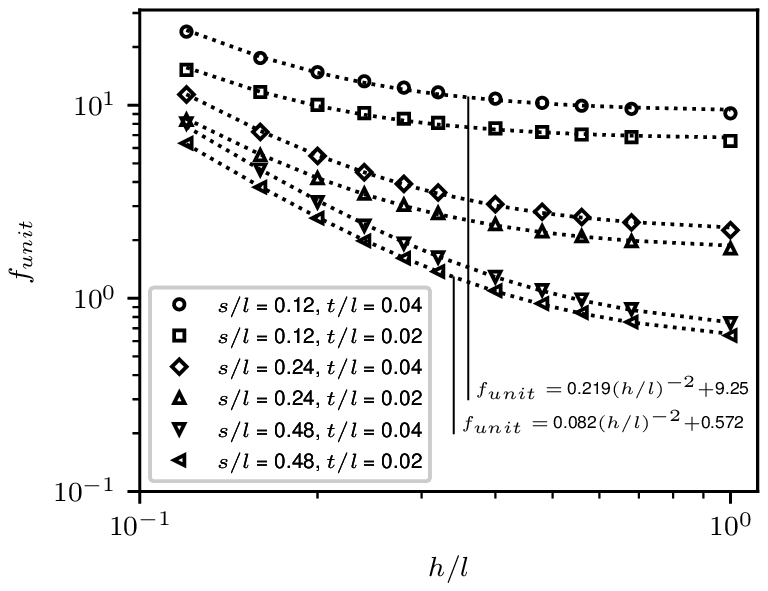}
\caption{\label{fig:HdataRE100} Influence of the fin height-to-length ratio on the friction factor for steady periodically developed flow, when $Re_{l}=100$ \newline\newline}
\end{minipage}
\hfill
\begin{minipage}{.475\textwidth}
% \centering
\raggedright
% \raggedleft
\includegraphics[scale = 1.00]{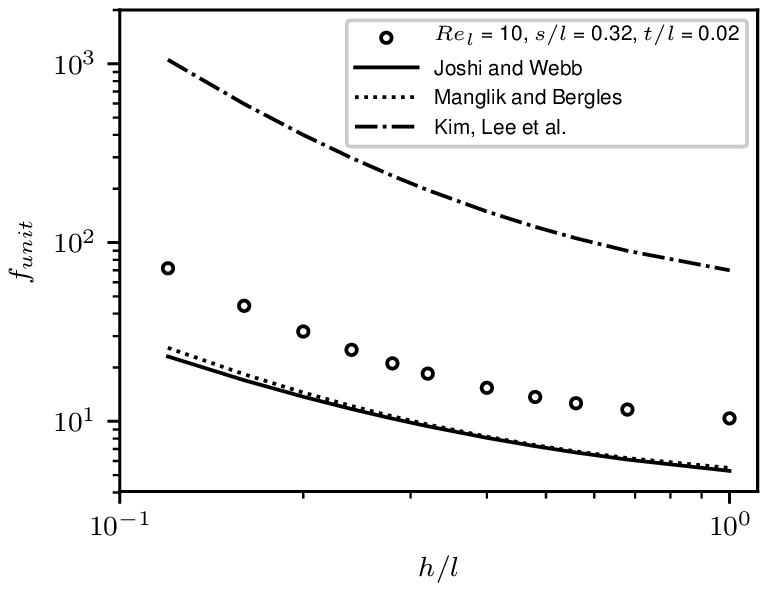}
\caption{\label{fig:litcompR10T2S32_nofit} A comparison between the friction factor correlations from the literature with the data from this work for steady periodically developed flow at small fin heights, when $Re_{l}=10$, $t/l=0.02$, $s/l=0.32$}
\end{minipage}
\end{figure}

From the data in Figure \ref{fig:HdataRE100}, the influence of the fin height-to-length ratio $h/l$ on the friction factor in the steady periodically developed flow regime can be observed. 
For low values of $h/l$, this friction factor becomes proportional to $(h/l)^{-2}$ according to our data. 
This is analogous to the friction factor for developed flow between two parallel plates (Refs.~\onlinecite{sutera1993history}), as the viscous forces on the top and bottom plate become the main contribution to the pressure drop for fins of a small height. 
On the contrary, for high values of $h/l$, the friction factor becomes independent of the fin height-to-length ratio. 
This can be explained by the fact that, as the relative fin height $h/l$ increases, the area of the fin sides increases, while the area of the top and bottom plate in contact with the flow remains the same. 
Therefore, the total pressure drop will mainly originate from the total drag (i.e. viscous drag and form drag) at the fin sides, when $h/l$ increases. 
Since the area of the fin sides depends on the relative fin pitch $s/l$, the relative fin thickness $t/l$ of the fins, as well as the relative fin height $h/l$, one would expect $h/l$ to still influence the friction factor. 
However, when $h/l$ increases, the flow field around the fin sides also becomes more two-dimensional, as its variation along the direction $\textbf{e}_{2}$ diminishes, so that eventually the drag at the fin sides is no longer affected by the relative fin height $h/l$. 

Based on the discussed trends, the dependence of the friction factor on the height-to-length ratio may be captured through a correlation of the form 
\begin{equation}
\label{eq: correlation form h/l}
f_{unit} \simeq A \left(\frac{h}{l}\right)^{-2} + B,
\end{equation}
where the parameters A and B depend on the Reynolds number $Re_l$, as well as the remaining geometrical parameters. 
The latter correlation form shows a good qualitative agreement with our data, when the parameters $A$ and $B$ are determined through a least-squares fitting.
As one can see in Figure \ref{fig:HdataRE100}, the correlation form (\ref{eq: correlation form h/l}) has an average and maximum relative error of 1\% and 6\%, respectively. 

Figure \ref{fig:HdataRE100} also shows that the shift between the two asymptotic trends $f_{unit} \sim (h/l)^{-2} $ and $f_{unit} \sim (h/l)^{0}$ is significantly influenced by the fin pitch-to-length ratio $s/l$, because the aspect ratio $s/h$ dictates the relative contribution of the friction on the top and bottom plate to the overall friction factor. 
Therefore, the trend $f_{unit} \sim (h/l)^{0}$ is more dominant when $s/l$ becomes smaller. 

In the literature, the influence of the fin height relative to the fin length has been incorporated in the friction factor both in an implicit and explicit way. 
Implicitly, the fin height-to-length ratio $h/l$ affects the Reynolds number $Re$, as the fin height $h$ appears in the expression for the hydraulic diameter. 
Explicitly, the influence of the fin height-to-length ratio has been included in the friction factor correlation through a factor $(h/l)^{e}$ such that $f_{unit} \sim (h/l)^{e} g(Re)$, where $g(Re)$ is a function of the Reynolds number $Re$ and the exponent $e$ lies between 0.02 and 2. 
In the correlations based on the experimental data of Kays and London (Refs.~\onlinecite{wieting1975empirical,joshi1987heat,manglik1995heat}), the resulting effective exponent $E$, such that $f_{unit} \sim (h/l)^E$, lies between -1.3 and -1.5 for low fin height-to-length ratios. 

Hence, it is clear that the correlations from the literature, which have been obtained for the larger $h/l$ ratios common in conventional channels, do not respect the correct trend for the small $h/l$ values that are encountered in micro- and mini-channels. 
As a consequence, the accuracy of the correlations from the literature rapidly diminishes when $h/l$ decreases because small $h/l$ values tend to fall outside the validity range of these correlations. 
In Figure \ref{fig:litcompR10T2S32_nofit} it is shown that the incorrect scaling of the correlations from the literature with $h/l$ results in an underestimation of our friction factor data in the laminar regime. 
Even if the correlations from the literature are rescaled with a constant to minimize the difference with our data and to account for incorrect scaling for other parameters, we still observe an average relative error of 30\% and a maximum relative error up to 65\% for $h/l$ near 0.1. 

In addition, it should be noted that from the more recent numerically determined correlation of Kim, Lee, et al. (Refs.~\onlinecite{kim2011correlations}), no continuous scaling laws for the variation of $f_{unit}$ with $h/l$ could be derived, since the latter correlation depends on the fin porosity, and thus also the fin height-to-length ratio, in a discontinuous way.
Finally, we remark that for all correlations from the literature, the friction factor still increases with $h/l$ for large fin height-to-length ratios, which is contradicted by our observation that $f_{unit}$ becomes independent of $h/l$ in that case. \\

% \newpage
% \clearpage

\subsection{\label{sec:influence s/l}The influence of the fin pitch-to-length ratio $s/l$ on the friction factor}

\begin{figure}[ht]
\centering
\begin{minipage}{.475\textwidth}
% \centering
\raggedleft
\includegraphics[scale = 1.00]{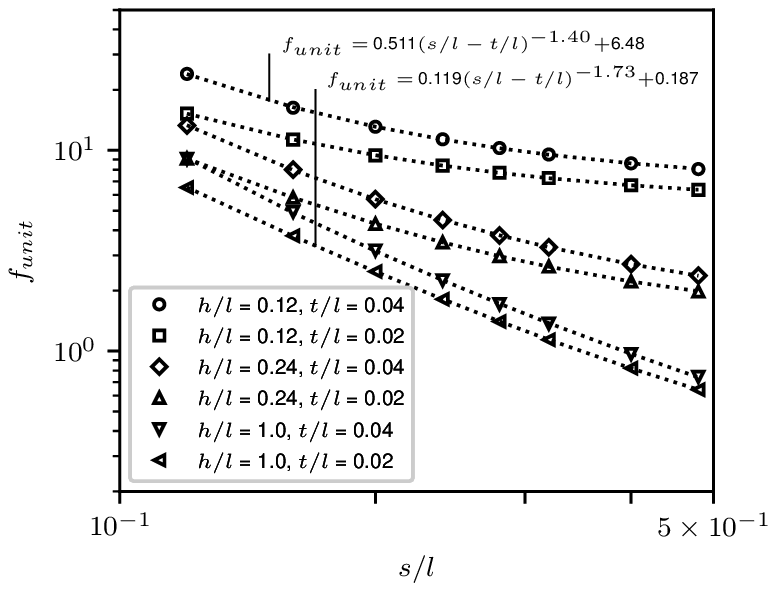}
\caption{\label{fig:SdataR100} Influence of the fin pitch-to-length ratio on the friction factor for steady periodically developed flow, when $Re_{l}=100$ \newline\newline}
\end{minipage}
\hfill
\begin{minipage}{.475\textwidth}
% \centering
\raggedright
% \raggedleft
\includegraphics[scale = 1.00]{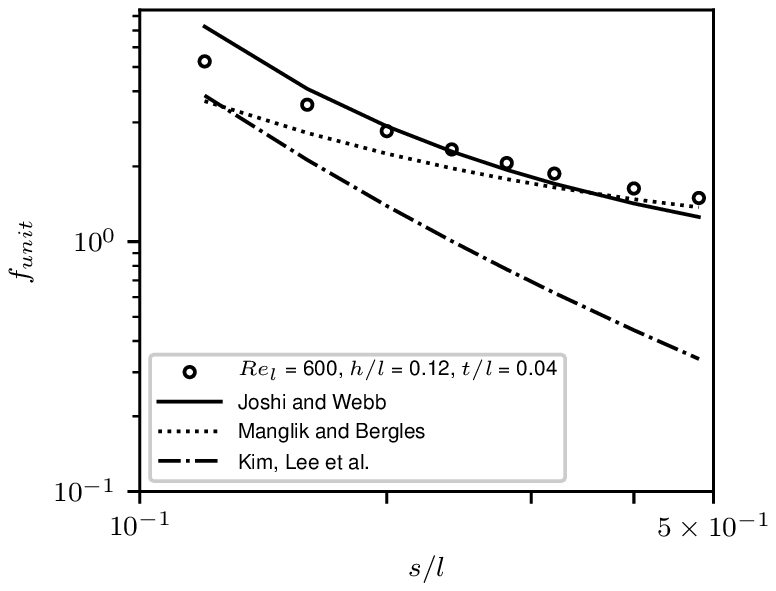}
\caption{\label{fig:litcompR600T4H12_nofit} A comparison between the friction factor correlations from the literature with the data from this work for steady periodically developed flow at small fin heights, when $Re_{l}=600$, $t/l=0.04$, $h/l=0.12$}
\end{minipage}
\end{figure}

Figure \ref{fig:SdataR100} displays the influence of the fin pitch-to-length ratio $s/l$ on the friction factor in the steady periodically developed flow regime. 
It can be seen that as $s/l$ approaches the fin thickness $t/l$, the friction factor approaches infinity because the channel becomes blocked. 
The influence of the fin pitch-to-length ratio $s/l$ on $f_{unit}$ is therefore best analyzed via the factor $(s-t)/l$.
As illustrated in Figure \ref{fig:SdataR100}, the friction factor is quite accurately expressed as a function of $(s-t)/l$ by correlations of the form 
\begin{equation}
\label{eq: correlation form s/l}
f_{unit} \simeq A\left(\frac{s}{l} - \frac{t}{l}\right)^{k} + B \,,
\end{equation} 
where $A$ and $B$ depend on the Reynolds number $Re_l$, as well as $h/l$ and $t/l$.
The negative exponent $k$, which ensures that $f_{unit} \rightarrow \infty$ for $s\rightarrow t$, can be taken as approximately constant over a wide range of Reynolds numbers, and over relatively broad intervals of $h/l$ and $t/l$. 
For instance, for $Re_l \in (1,600)$, $t/l \in (0.01, 0.02) $ and $h/l \in (0.5, 1) $, the previous correlation form holds within a relative error of 5\% when $k=-1.69$. 
Yet, the exponent $k$ in (\ref{eq: correlation form s/l}) is not a constant over the entire set of our friction factor data. 

The proposed correlation form (\ref{eq: correlation form s/l}) predicts that for high values of $s/l$, the friction factor becomes independent of the fin pitch-to-length ratio $s/l$.
This is in agreement with our data, as it can be seen in Figure \ref{fig:SdataR100}. 
The reason is that in that case, the overall pressure gradient stems primarily from the friction forces on the top and bottom plate of the channel, which are primarily dependent on the fin height-to-length ratio, so that $f_{unit} \sim (h/l)^{-2}$ as discussed in the previous subsection. 
As this trend originates from a reduced contribution of the total drag at the fin sides to the pressure gradient, its manifestation is greatly influenced by the relative fin height $h/l$, which is illustrated in Figure \ref{fig:SdataR100} for a few geometries. 
More specifically, for smaller values of $h/l$, the trend $f_{unit} \sim (s/l-t/l)^{0}$ will become more dominant. 

Similar to the fin height-to-length ratio, the influence of the fin pitch-to-length ratio on the friction factor is usually incorporated in the correlations from the literature through a formula of the form $f \sim (s/l)^{k}$ for some negative exponent $k$. 
Of the correlations listed in Tables \ref{tab:literature} and \ref{tab:literature2}, Joshi and Webb's correlation  (Refs.~\onlinecite{joshi1987heat}) is the only one which accounts for the limit that the friction factor becomes infinite when $s-t$ approaches zero. 
This limit, however, is incorporated implicitly in the definition of the friction factor, via a hydraulic diameter $D_{h} \sim (s-t)$ and a reference velocity $U_{ref} \sim (s-t)^{-1}$, which give rise to an infinitely large pressure drop when the fin pitch $s$ approaches the fin thickness $t$. 
% which reduces the Reynolds number $Re_{D_h}$ to zero for the fin pitch $s$ going to the fin thickness $t$. 
% , despite the porosity is still non-zero in that case
On the other hand, for small fin pitch-to-length ratios, i.e. $s/l < 0.25$, the correlation of Joshi and Webb (Refs.~\onlinecite{joshi1987heat}) overestimates the friction factor by 5\% to 75\%.
Also the other correlations deviate within the same order of magnitude, albeit they typically show an underestimation, which is as much as 75\% for $s/l>0.4$. 
This is illustrated in Figure \ref{fig:litcompR600T4H12_nofit}. 
Again, these errors have been determined after rescaling the correlations from the literature with a constant, to minimize the deviation from our data and compensate for the wrong scaling with other parameters. 
For larger values of $s/l$, the friction factor still depends on the fin pitch-to-length ratio according to all correlations from Tables \ref{tab:literature} and \ref{tab:literature2}. 
Most notably, this is the case for the correlation of Kim, Lee et al. (Refs.~\onlinecite{kim2011correlations}), as shown in Figure \ref{fig:litcompR600T4H12_nofit}.
Their correlation significantly underestimates the friction factor up to 70\% for $s/l$ near 0.5. 
Furthermore, this correlation predicts a discontinuous dependence of the friction factor on the fin pitch-to-length ratio due to its piecewise dependence on the fin porosity. \\

% \newpage
% \clearpage

\subsection{\label{sec:influence t/l}The influence of the fin thickness-to-length ratio $t/l$ on the friction factor}

\begin{figure}[ht]
\centering
\begin{minipage}{.475\textwidth}
% \centering
\raggedleft
\includegraphics[scale = 1.00]{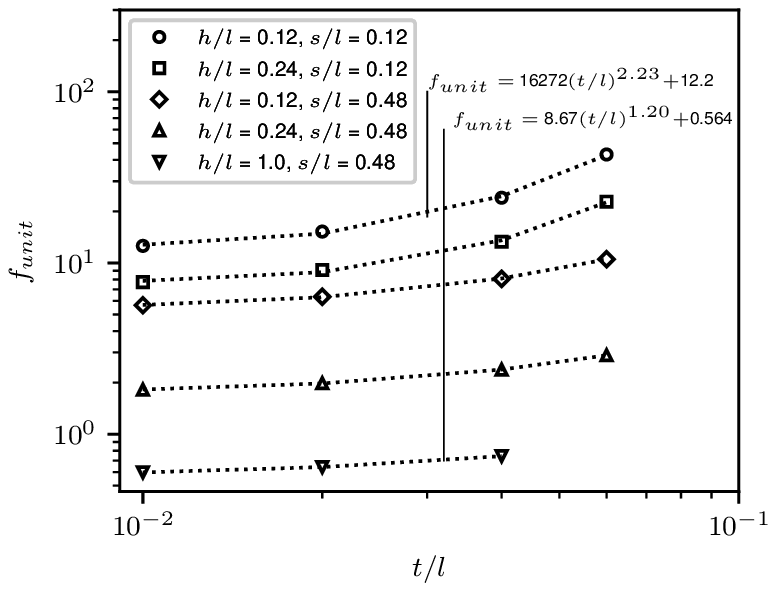}
\caption{\label{fig:TdataR100} Influence of the fin thickness-to-length ratio on the friction factor for steady periodically developed flow, when $Re_{l}=100$ \newline\newline}
\end{minipage}
\hfill
\begin{minipage}{.475\textwidth}
% \centering
\raggedright
% \raggedleft
\includegraphics[scale = 1.00]{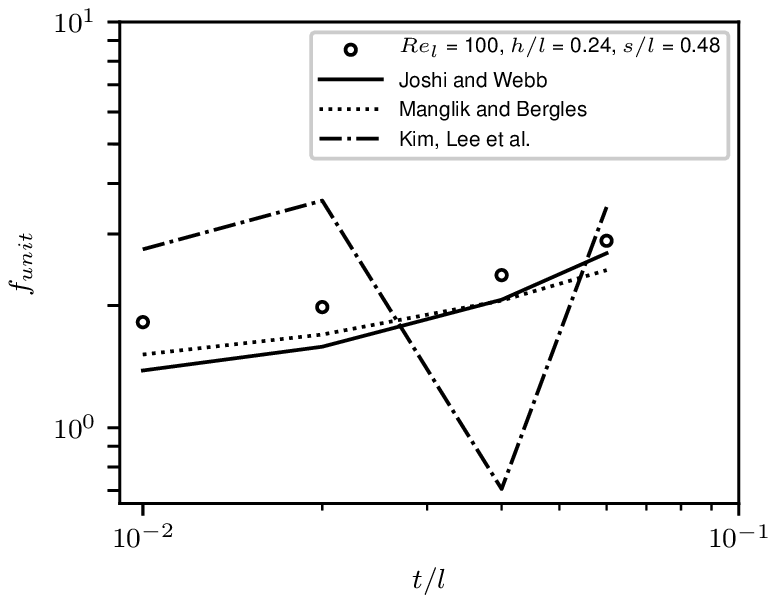}
\caption{\label{fig:litcompR100H24S48_nofit} A comparison between the friction factor correlations from the literature with the data from this work for steady periodically developed flow at small fin heights, when $Re_{l}=100$, $h/l=0.24$, $s/l=0.48$}
\end{minipage}
\end{figure}

The final geometrical parameter is the fin thickness-to-length ratio $t/l$, whose influence on the friction factor is shown in Figure \ref{fig:TdataR100}. 
In this figure, it can be seen that when the fin thickness-to-length ratio $t/l$ goes to zero, the friction factor becomes independent of $t/l$. 
This can be explained from the fact that the form drag over the fin sides, which are normal to the main flow direction $\boldsymbol{e}_1$ and which give the fin its thickness $t$, mainly results from the presence of the wakes behind the fin (see Figure \ref{fig:streamlinesRE100_400T6H48S12}).
So, for small fin thicknesses, these wakes become small as well, and the contribution of the corresponding form drag force on $f_{unit}$ becomes negligible. 
This has been observed by quantifying the contribution of the viscous and form drag to the overall pressure gradient for various fin thicknesses. 
Obviously, Figure \ref{fig:TdataR100} demonstrates again that when $t/l$ approaches the fin pitch-to-length ratio $s/l$, the friction factor increases to infinity, as described already in the previous subsection. 

A closer inspection of our friction factor data has revealed that, in good approximation, the friction factor depends on $t/l$ through a relationship of the form
\begin{equation}
\label{eq: correlation form t/l}
f_{unit} \simeq A\left(\frac{t}{l} \right)^n +B\,.
\end{equation}
Here, $n$ is a positive constant, while $A$ and $B$ are assumed to represent functions of the Reynolds number $Re_l$, as well as $h/l$ and $s/l$, or $(s-t)/l$.
In Figure \ref{fig:TdataR100}, curves for $f_{unit}$ versus $t/l$ have been fitted according to the form (\ref{eq: correlation form t/l}). 
% to illustrate that that this form does not deviate more than $xx\%$ from our data, if the exponent $n$ equals $XX$. 
It can be observed that the positive exponent $n$ can not be considered a constant over all the friction factor data. 
However, $n$ can be approximated by a constant over separate ranges of the Reynolds number $Re_l$. 
For example, for $Re_l \in (1,200)$ and all values of $s/l$ and $h/l$ included in this work, the correlation form (\ref{eq: correlation form t/l}) results in a relative error below 5\% with our data when $n=0.84$. 

In order that the proposed form (\ref{eq: correlation form t/l}) for the friction factor respects the asymptotic trend that $f_{unit}$ becomes independent of $t/l$ for $t/l \rightarrow 0$, $B$ can only be a function $s/l$, but not $(s-t)/l$. 
On the other hand, the form (\ref{eq: correlation form t/l}) is only compatible with the previously proposed form (\ref{eq: correlation form s/l}) for the dependence of $f_{unit}$ on $(s-t)/l$, if $A \sim (s/l-t/l)^{e}$ with $e<0$. 

%This means that any correlation of the form $f \sim (s/l-t/l)^{e} g(t/l)$, as common in the literature, can capture the trends for $t/l \rightarrow s/l$ and $t/l \rightarrow 0$, only if the exponent $e$ satisfies $e <0$, and the function $g$ satisfies $g(0)=A$ with $A$ some constant.
%Nevertheless, virtually all correlations from the literature, have form based on $e=0$ and $g(t/l)=(t/l)^n$

For virtually all friction factor correlations, the latter restrictions on the form (\ref{eq: correlation form t/l}) are violated, as these correlations are based on the assumption that $f_{unit} \sim (t/l)^n$.
Only the correlation of Joshi and Webb (Refs.~\onlinecite{joshi1987heat}) accounts for the observation that friction factor becomes independent of $t/l$ for small fin thickness values.
All other correlations listed in Tables \ref{tab:literature} and \ref{tab:literature2} predict an asymptotic behaviour of the form $f_{unit} \sim (t/l)^{n}$ with $n$ between 0.03-0.3 for $t/l \rightarrow 0$. 
This results in an underestimation of the friction factor by approximately 20\% for $t/l<0.02$, after rescaling the correlation with a constant to exclude the wrong scaling with other parameters. 
For the correlation of Kim, Lee et al. (Refs.~\onlinecite{kim2011correlations}), the friction factor does not depend on the fin thickness-to-length ratio in a continuous manner, since its mathematical relationship to $t/l$ is expressed through a discontinuous piecewise function of the porosity, as depicted in Figure \ref{fig:litcompR100H24S48_nofit}. 

%\newpage
%\clearpage

\subsection{\label{sec:critical} The influence of the critical Reynolds number}

Although the correlation form (\ref{eqn:ergun}) accurately captures our friction factor data for periodically developed flow in offset strip fin arrays in micro- and mini-channels, it is actually an approximation from a physical point of view.
The reason is that the assumption that the Forchheimer correction term scales quadratically with the magnitude of the volume-averaged velocity only holds for the strong inertia regime, as discussed in Subsection \ref{sec:influence re}.
Nevertheless, the following more detailed analysis of our friction factor data reveals that also a so-called weak inertia regime  (Refs.~\onlinecite{mei1991effect,koch1997moderate,amaral1997dispersion,lasseux2011stationary}) occurs in offset strip fin arrays at very low Reynolds numbers.
In the weak inertia regime, the Forchheimer correction term is known to display a cubic dependence on the magnitude of the volume-averaged velocity, so that $f_{unit} = A Re_l^{-1} + C Re_l $ instead of $f_{unit} = A Re_l^{-1} + B $.

\begin{figure}[ht]
\includegraphics[scale = 1.00]{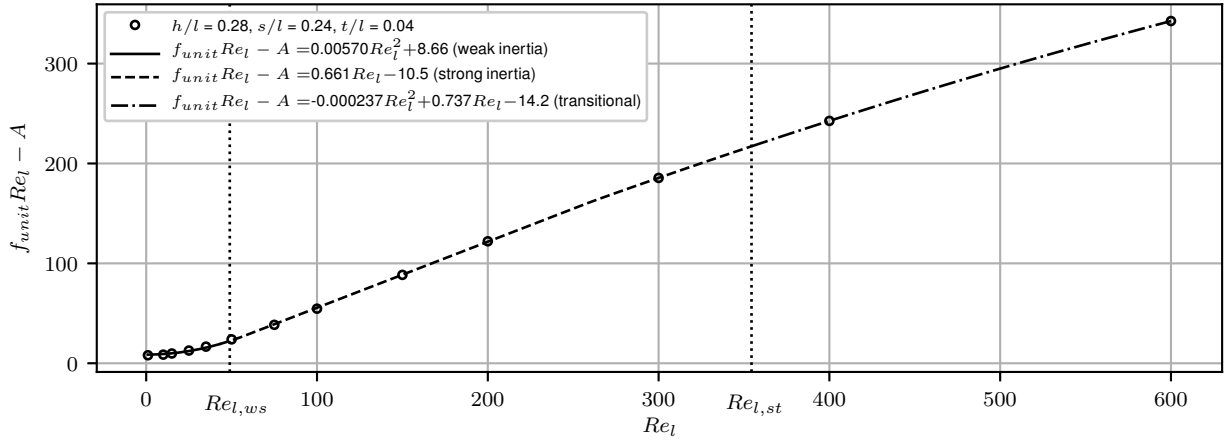}
\caption{\label{fig:f_correction_fit} Distinction between the flow regimes based on the Forchheimer correction term for steady periodically developed flow, when $h/l = 0.28$, $s/l = 0.24$, $t/l = 0.04$}
\end{figure}

In Figure \ref{fig:f_correction_fit}, the existence of both a strong and weak inertia regime has been visualized by plotting the Forchheimer correction term $\left( f_{unit} Re_{l} - A \right)$ with respect to the Reynolds number $Re_{l}$.
To this end, the parameter $A$ was determined for each unit cell geometry by fitting the relation $f_{unit} = A Re_l^{-1}$ through the friction factor data points in the lower Reynolds number range. 
This range is determined by the values of $Re_{l}$ where $f_{unit}$ can be approximated by $A Re_l^{-1}$ within an accuracy of 1\%, for some constant value of $A$. 
% For all geometrical parameters, this corresponds to the friction factor data points at the lowest Reynolds numbers included in this study. 

The strong inertia regime, which justifies the theoretical validity of (\ref{eqn:ergun}), extends over the Reynolds number range $Re_l \in \left(Re_{l,ws} , Re_{l,st}\right)$
in Figure \ref{fig:f_correction_fit}.
The two critical Reynolds numbers $Re_{l,ws}$ and $Re_{l,st}$ mark the transition from the weak inertia regime to the strong inertia regime and the transition from the strong inertia regime to the transitional regime, respectively. 
Both correspond to the values of $Re_l$ at which the correction term $\left( f_{unit} Re_{l} - A \right)$ deviates 5\% from the linear dependence $\left( f_{unit} Re_{l} - A \right) \simeq B Re_{l} + C$, as in the study (Ref.~\onlinecite{lasseux2011stationary}). 
In the strong inertia regime, the constant $C$ is found to be small with respect to $A$ for all the geometries in this work.
This confirms the expected purely quadratic dependence of the Forchheimer correction term on the magnitude of the volume-averaged velocity in this regime. 

The weak inertia regime in Figure \ref{fig:f_correction_fit}, which is characterized by $\left( f_{unit} Re_{l} - A \right) \simeq D Re_l^2 + F$, is found for $Re_l \in \left(0, Re_{l,ws}\right)$. 
Also, the constant $F$ remains small relative to $A$, which confirms a pure cubic relation between the Forchheimer correction term and $\Vert\langle \boldsymbol{u} \rangle\Vert$ in this regime. 
Furthermore, at higher Reynolds numbers, $Re_l > Re_{l,st}$, a transitional flow regime can be recognized in Figure \ref{fig:f_correction_fit}. 
For Reynolds numbers in the transitional flow regime, the periodically developed flow equations (\ref{eq:NavierStokes}) have time-dependent, chaotic flow solutions on a unit cell. 
Therefore, the friction factor in the transitional regime is actually defined based on the ensemble-averaged pressure gradient $\Vert \overline{\nabla \mathrm{P}} \Vert$ instead of $\Vert \nabla \mathrm{P} \Vert$, as a constant $\langle \boldsymbol{u} \rangle$ was imposed.
% Therefore, the Reynolds numbers in the transitional regime are actually defined based on the ensemble-and-volume-averaged velocity $\langle \bar{\boldsymbol{u}} \rangle$ instead of $\langle \boldsymbol{u} \rangle$.
Nevertheless, it still remains a question whether these chaotic flow solutions are representative of the complete flow through a finite fin array in a channel (Ref.~\onlinecite{hill2002moderate}). 
In any case, our friction factor data indicates that in the transitional regime, the correction $\left( f_{unit} Re_{l} - A \right)$ can be captured by a quadratic dependence of the Reynolds number, comparable to the weak inertia regime.
This is in agreement with the study of Lasseux et al. (Ref.~\onlinecite{lasseux2011stationary}).

\begin{figure}[h!]
\includegraphics[scale = 1.00]{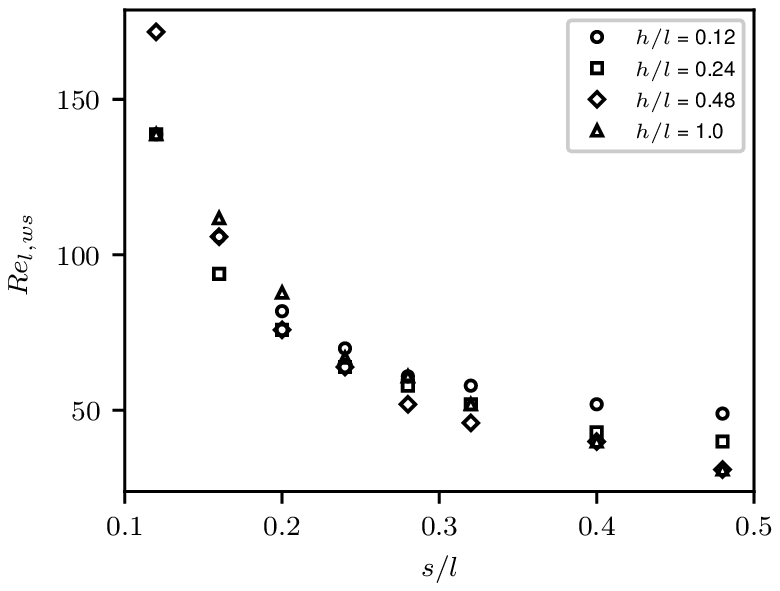}
\caption{\label{fig:ReCrit_weak} Critical Reynolds number representing the transition of the weak inertia regime towards the strong inertia regime, when $t/l = 0.02$}
\end{figure}

Figure \ref{fig:ReCrit_weak} depicts the critical Reynolds number $Re_{l,ws}$ as a function of the fin pitch-to-length ratio $s/l$.
A monotonous decrease of $Re_{l,ws}$ can be observed for increasing $s/l$. 
This figure also shows that no significant influence of the fin height-to-length ratio $h/l$ and fin thickness-to-length ratio $t/l$ on the critical Reynolds number $Re_{l,ws}$ has been found. 

\begin{figure}[h!]
\centering
\begin{minipage}{.475\textwidth}
% \centering
\raggedleft
\includegraphics[scale = 1.00]{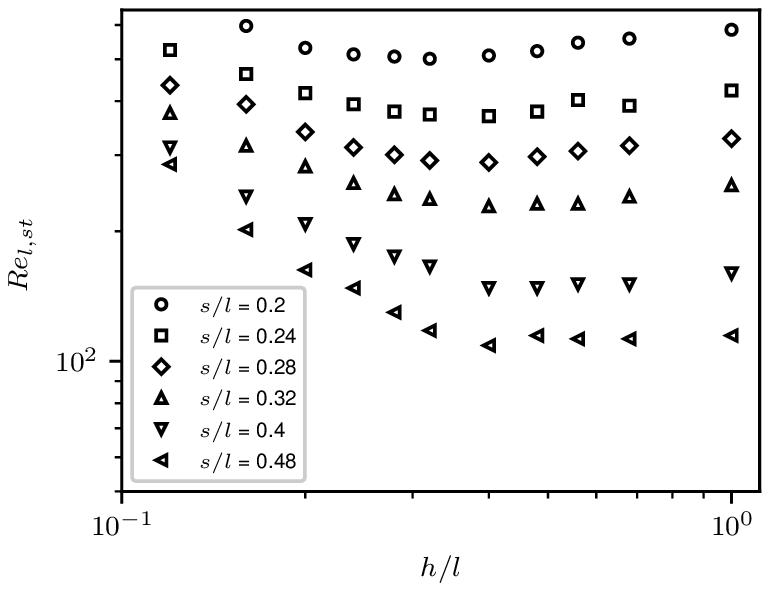}
\caption{\label{fig:ReCrit_strongH} Critical Reynolds number representing the transition of the weak inertia regime towards the strong inertia regime, when $t/l = 0.02$}
\end{minipage}
\hfill
\begin{minipage}{.475\textwidth}
% \centering
\raggedright
% \raggedleft
\includegraphics[scale = 1.00]{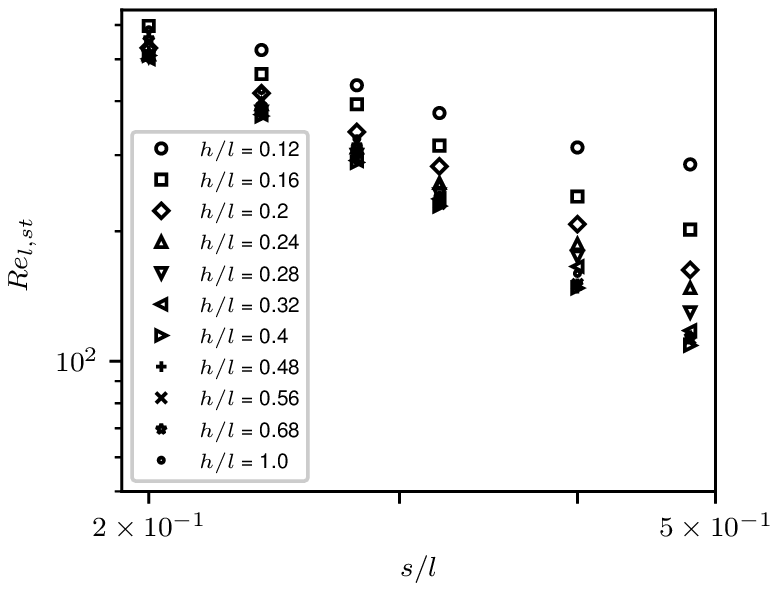}
\caption{\label{fig:ReCrit_strongS} Critical Reynolds number representing the transition of the strong inertia regime towards the transitional regime, when $t/l = 0.02$}
\end{minipage}
\end{figure}

The critical Reynolds number $Re_{l,st}$ is shown in Figures \ref{fig:ReCrit_strongH} and \ref{fig:ReCrit_strongS}. 
It can be seen that for a higher value of the fin height-to-length ratio $h/l$, transition tends to occur at a lower Reynolds number. 
However, as illustrated in Figure \ref{fig:ReCrit_strongH}, for fin height-to-length ratios larger than 0.4, $Re_{l,st}$ becomes independent of $h/l$. 
This can again be observed in Figure \ref{fig:ReCrit_strongS}, which also shows that the fin pitch-to-length ratio $s/l$ has an inverse relationship with $Re_{l,st}$. 
Finally, we add that the fin thickness-to-length ratio $t/l$ has been found to have no significant influence on the critical Reynolds number $Re_{l,st}$. 

\newpage
\clearpage


The previous figures indicate the Reynolds number range over which a direct proportionality between the correction $\left( f_{unit} Re_{l} - A \right)$ and $B Re_{l}$ is valid from a physical perspective. 
However, from a practical perspective, the correlation form (\ref{eqn:ergun}) is able to accurately capture the periodically developed friction factor data for offset strip fin arrays in micro- and mini-channels.

% \newpage
% \clearpage

\section{\label{sec:correlation}Friction factor correlation}

\subsection{Fitting approach}

To obtain an accurate friction factor correlation for steady periodically developed flow through offset strip fin arrays in micro- and mini-channels, several heuristically constructed candidate correlations, all respecting the forms (\ref{eqn:ergun})-(\ref{eq: correlation form t/l}), were fitted to the friction factor data from this work. 
%An overview of these correlation models is included in Appendix \ref{sec:bayes}.
For the fitting procedure, two methods were applied. 
First, a non-linear least-squares optimization method based on the trust region reflective algorithm (Refs.~\onlinecite{2020SciPy-NMeth}) was used to find the \textit{optimal} parameter values for each candidate correlation. 
Secondly, the Bayesian approach for parameter estimation and model validation (Refs.~\onlinecite{de2021bayesian}) was used to find the \textit{most likely} parameter values for each candidate correlation.
The Bayesian approach was also employed to quantify the so-called \textit{log-evidence} for each candidate correlation, which is a statistical measure for its ability to capture the data.
Finally, the candidate correlation and the fitting parameters leading to the highest relative accuracy and highest log-evidence with respect to our friction factor data were selected.
We remark that for this final correlation, which is discussed in the next subsection, the \textit{optimal} and \textit{most likely} parameter values were found to be equal within their significant digits. 

For an explanation on how the most likely parameters and log-evidence of each candidate correlation were quantified via the Bayesian approach, we refer in the first place to the relevant literature as e.g. in (Refs.~\onlinecite{de2021bayesian}). 
For the sake of transparency, we do clarify that, given a certain correlation model $\mathcal{M}$, the Bayesian approach allows for the computation of the joint probability density function of the fitting parameters $\boldsymbol{\theta}$, and thus their most likely value, having observed a given data set $\mathcal{D}$ of friction factor values. 
This joint probability density function, or so-called posterior, has been calculated as
\begin{equation}
    \mathcal{P} \left( \boldsymbol{\theta} | \mathcal{D}, \mathcal{M} \right) = \frac{\mathcal{L} \left( \mathcal{D} | \boldsymbol{\theta}, \mathcal{M} \right) \mathcal{P} \left( \boldsymbol{\theta} | \mathcal{M} \right)}{\mathcal{P} \left( \mathcal{D} | \mathcal{M} \right)}. 
\end{equation}
In this equation, the likelihood $\mathcal{L} \left( \mathcal{D} | \boldsymbol{\theta}, \mathcal{M} \right)$ is the probability density function of observing the data $\mathcal{D}$ for the given fitting parameters $\boldsymbol{\theta}$.
It thus represents the correlation error, which we assume to have distribution equal to the multiplication of different Gaussian distributions for every data point $\mathcal{D}_{i}$, similar to (Refs.~\onlinecite{de2021bayesian}), each with a mean of 0 and a standard deviation of $0.01 \mathcal{D}_{i}$. 
The prior $\mathcal{P} \left( \boldsymbol{\theta} | \mathcal{M} \right)$ is the probability density function of the fitting parameters $\boldsymbol{\theta}$ prior to observing the data $\mathcal{D}$ and depends on the mathematical constraints of the correlation model. 
For our study, this probability density function for each fitting parameter $\boldsymbol{\theta}_{i}$ has been assumed to be uniform since any a priori information regarding the fitting parameters is absent. 
Finally, the evidence $\mathcal{P} \left( \mathcal{D} | \mathcal{M} \right)$ is the probability to observe the data $\mathcal{D}$ regardless of the fitting parameter values $\boldsymbol{\theta}$ and is computed through the expression $\mathcal{P} \left( \mathcal{D} | \mathcal{M} \right) = \int \mathcal{L} \left( \mathcal{D} | \boldsymbol{\theta}, \mathcal{M} \right) \mathcal{P} \left( \boldsymbol{\theta} | \mathcal{M} \right) \,d\boldsymbol{\theta}$. 
The log-evidence $\log \left[ \mathcal{P} \left( \mathcal{D} | \mathcal{M} \right) \right]$ thus provides a quantitative measure for the suitability of the correlation model $\mathcal{M}$ to represent the given data. 
Therefore, it has been used to compare the different candidate correlations models for the friction factor. 
As a last remark, we add that the Laplace approximation was introduced for the computation of the evidence, which means that the posterior is assumed to have a Gaussian distribution. 
This is true for a large number of data samples of independent and identically distributed fitting parameters (Refs.~\onlinecite{de2021bayesian}), as it is the case in this study. 

%\newpage
%\clearpage

\subsection{Fitting result}
By evaluating several heuristically constructed candidate correlations through the former fitting procedures, the following correlation for the friction factor in the steady periodically developed flow regime was obtained: 
\begin{equation}
    f_{unit} = c_{0} Re_l^{-1} + c_{1}, 
% \label{eq:fit_aligned}
\nonumber
\end{equation}
with
\begin{equation}
\begin{aligned}
  c_{0} &= [23.5 (s/l-t/l)^{-0.83} + 14.9](t/l)^{0.84} (h/l)^{-2} \\
        & \qquad + 13.0 (s/l-t/l)^{-1.69} + 6.0 (h/l)^{-2}, \\
  c_{1} &=  56.5  (s/l-t/l)^{-1.34} (t/l)^{2.94} (h/l)^{-1.08} \\
        & \qquad + 0.0355(s/l-t/l)^{-0.83}.\\
\end{aligned}
\label{eq:fit_aligned2}
\end{equation}
For the friction factor data in this work, this final correlation results in an average relative error of 2\%.
Its maximal relative error is below 4\%, 5\% and 8\% for respectively 90\%, 95\%, 99\% of the data points. 
Besides having the highest accuracy, the correlation form (\ref{eq:fit_aligned2}) resulted in the largest log-evidence value of all the considered candidate correlations. 
For example, regarding the dependence of the friction factor on the Reynolds number, the form $f_{unit} = c_{0} Re_l^{-1} + c_{1}$ resulted in a log-evidence values 10 times larger than that of the form $f_{unit} = c_{0} Re_l^{e}$, which is typically used in the literature (see Tables \ref{tab:literature} and \ref{tab:literature2}). 
Furthermore, through the joint probability density function obtained from the Bayesian approach, the standard deviations of all the exponents in expression (\ref{eq:fit_aligned2}) are observed to remain below 1\% of the exponent value. 
This confirms the uniqueness of all the exponent values and the prevention of over-fitting the data, which was further reflected in the high log-evidence value of the presented correlation form. 
% A more detailed discussion of the results from the Bayesian study can be found in REFERENCE ARXIV DOCUMENT (upload first draft, which is ready, and update if needed).

\begin{figure}[ht]
\centering
\begin{minipage}{.475\textwidth}
% \centering
\raggedleft
\includegraphics[scale = 1.00]{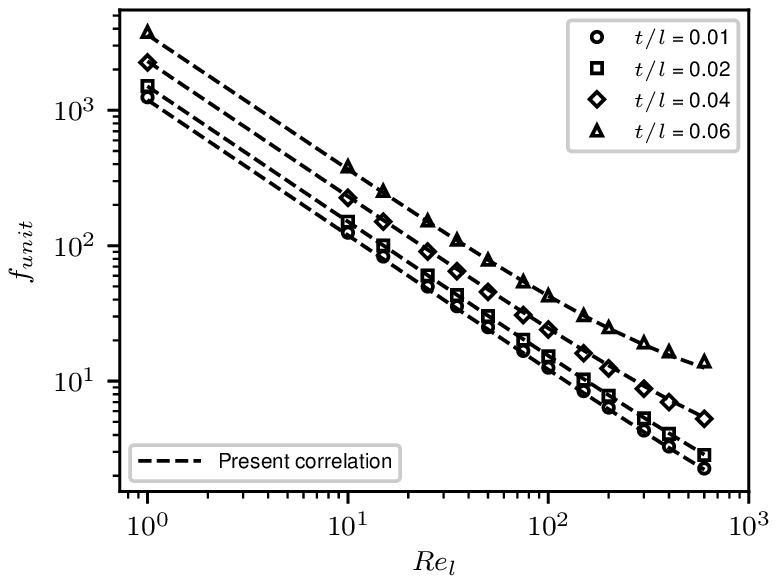}
\caption{\label{fig:fitREdataH12S12} A comparison between the final friction factor correlation with data from this work for steady periodically developed flow at small fin heights, when $h/l=0.12$, $s/l=0.12$}
\end{minipage}
\hfill
\begin{minipage}{.475\textwidth}
% \centering
\raggedright
% \raggedleft
\includegraphics[scale = 1.00]{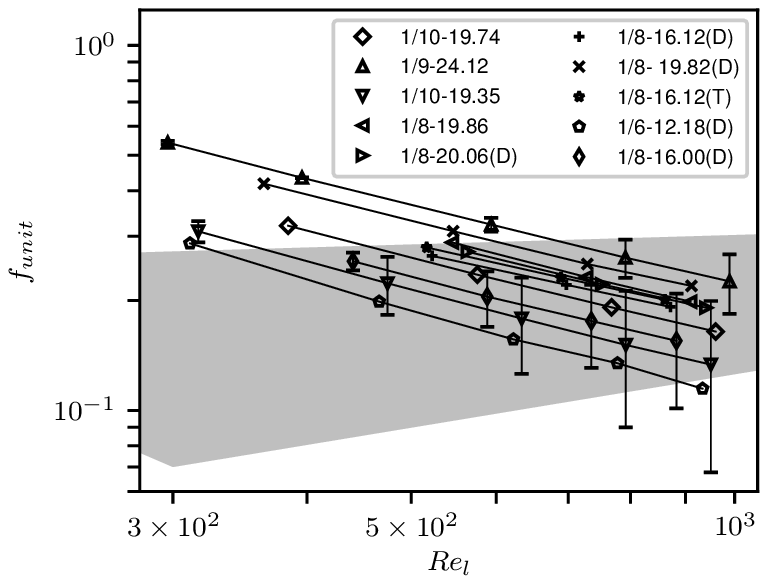}
\caption{\label{fig:litcompKLdata} Discrepancies between the final friction factor correlation and the data from Kays and London (Refs.~\onlinecite{kays1984compact}), where the grey zone indicates the region of transitional flow}
\end{minipage}
\end{figure}

As an illustration of the correlation's accuracy for the friction factor data from this work, it has been compared with a selection of data points in Figure \ref{fig:fitREdataH12S12}. 
In Figure \ref{fig:litcompKLdata}, the presented correlation (\ref{eq:fit_aligned2}) is compared with the experimental friction factor data from Kays and London (Refs.~\onlinecite{kays1984compact}). 
It can be seen that below a Reynolds number $Re_{l}$ of 600, the relative difference between our correlation and the data from Kays and London, indicated by the error bars, remains smaller than 10\%, and even decreases at lower $Re_{l}$, for all geometries with $h/l < 1$.
Taking into account that the experimental uncertainty on the friction factor and the Reynolds number from (Refs.~\onlinecite{kays1984compact}) are equal to 5\% and 2\% respectively, this relative difference falls well within our correlation's accuracy. 
We note that the grey zone in this last figure indicates again the expected region of transitional flow according to  (Refs.~\onlinecite{joshi1987heat,mochizuki1988flow,dejong1997experimental}). \\
% This is illustrated in Figure \ref{fig:litcompKLdata} for various offset strip fin geometries from (Refs.~\onlinecite{kays1984compact}). 
% Note that in this comparison, geometries with a relative fin height $h/l$ larger than 1 have also been included. \\

The correlation (\ref{eq:fit_aligned2}) is consistent with our observation that the assumption of a strong inertia regime (\ref{eqn:ergun}) is justified over almost the entire range of Reynolds numbers studied in this work, i.e. $Re_{l}=1-600$. 
Yet, it also incorporates all the trends (\ref{eqn:ergun})-(\ref{eq: correlation form t/l}) which describe the influence of the geometrical parameters, as discussed in Section \ref{sec:aligned}. 

Regarding the influence of the fin height-to-length ratio, the final correlation respects the asymptotic trends $f_{unit} \sim (h/l)^{-2}$ for $h/l \rightarrow 0$ and $f_{unit} \sim (h/l)^{0}$ for $h/l \rightarrow \infty$.
Due to the inclusion of the term $(h/l)^{-1.08}$ in the coefficient $c_{1}$, both of these asymptotic trends have been matched for intermediate values of $h/l$. 
Furthermore, in the first coefficient $c_{0}$, the term $6.0 (h/l)^{-2}$ corresponds to the friction factor for fully-developed flow between parallel plates: $f_{unit}(h/l) = 6[Re_{l}(h/l)]^{-1}$. 
This is considered the asymptotic value of the friction factor that is recovered when the fin height-to-length ratio $h/l$ and the fin thickness-to-length ratio $t/l$ approach zero. 

Regarding the influence of the fin pitch-to-length, it is clear that the limit $f_{unit} \rightarrow \infty$ for $s \rightarrow t$ has been incorporated as well. 
After all, the four different exponents for the term $(s/l-t/l)$, which have been included for an accurate fit over the entire range of $s/l$ values from this work, are all negative.
Moreover, in accordance with our observations, the correlation (\ref{eq:fit_aligned2}) becomes independent of the fin pitch-to-length ratio as $s/l$ increases. 

Finally, the presented relation for $f_{unit}$ also becomes independent of the fin thickness-to-length ratio as $t/l$ decreases. 
For larger values of $t/l$, two different exponents for the term $(t/l)$ account for the influence of the thickness-to-length ratio over different Reynolds number ranges. 
\vfill

% \begin{figure}[ht!]
% \includegraphics[scale =1.00]{figures/fitREdataH12S12.eps}
% \caption{\label{fig:fitREdataH12S12} Final friction factor correlation for steady periodically developed flow, when $h/l=0.12$, $s/l=0.12$}
% \end{figure}

\newpage
\clearpage

\section{\label{sec:conclusion}Conclusions}

The friction factor for steady laminar periodically developed flow through offset strip fins arrays in micro- and mini-channels has been examined in the present study.
The friction factor data was collected by solving numerically the periodically developed flow equations on a unit cell of the fin array. 
To this end, more than 2765 flow simulations have been carried out. 

It was found that the friction factor correlations in the literature generally cover applications of offset strip fins in larger conventional channels, so that very few data points in the laminar flow regime were considered. 
Therefore, they do not respect the correct asymptotic behaviour of the friction factor for low to moderate Reynolds numbers and result in deviations of 20\%-80\% and more for the laminar regime in micro- and mini-channels. 
Neither do they consider the asymptotic trends and limits observed for small and large values of the fin height, pitch and thickness relative to the fin length. 

To overcome these discrepancies, a new correlation has been presented for the periodically developed friction factor, whose maximal relative error is below 8\% with respect to our data. 
This new correlation was obtained through a least-square fitting procedure, though its suitability was further analyzed by means of the Bayesian approach for parameter estimation. 

The new correlation scales inversely linear with the Reynolds number, in agreement with our observation that a strong inertia regime prevails over nearly the entire range of Reynolds numbers up to 600. 
Nevertheless, a closer look at the dependence of the friction factor on the Reynolds number has revealed that also a regime of weak inertia occurs at low Reynolds numbers. In contrast, a transitional regime occurs at higher Reynolds numbers. 
The transitions from the strong inertia regime have been characterized by two critical Reynolds numbers in this work. 
Both critical Reynolds numbers are found to decrease monotonously with the fin pitch and to be independent of the fin thickness. 
Furthermore, the transition from the weak inertia regime to the strong inertia regime was shown to be independent of the fin height. 
Yet, the transition from the strong inertia regime towards the transitional regime occurs at lower Reynolds numbers for larger fin heights. 

Finally, the presented new correlation incorporates all the observed trends for variations in the geometrical parameters of the offset strip fin array. 
First, the correlation predicts that the friction factor becomes proportional to the inverse square of the fin height for small fin heights, analogous to fully-developed flow between two parallel plates.
Secondly, it predicts that the pressure gradient approaches infinity as the fin pitch becomes equal to the fin thickness. 
Additionally, the correlation respects the observation that the friction factor becomes independent of the fin height, fin pitch, and fin thickness for large fin heights, large fin pitches, and small fin thicknesses, respectively.

\section{\label{sec:contributions}Contributions}

The computational algorithms and software framework for the periodic flow equations were implemented and validated by G. Buckinx. 
All flow simulations and post-processing calculations
were performed by A. Vangeffelen. 
The interpretation of the results was done by A. Vangeffelen, with input from 
G. Buckinx regarding the existing literature. 
A. Vangeffelen and G. Buckinx wrote the paper with input from M. Baelmans and M. R. Vetrano.

\section{\label{sec:acknowledgements}Acknowledgements}
This work was partly funded by the Research Foundation — Flanders (FWO) through G. Buckinx's post-doctoral project grant 12Y2919N, and by the Flemish Institute for Technological Research (VITO) through A. Vangeffelen's Ph.D. grant 1810603. 

\newpage
\clearpage

\appendix

\section{\label{sec:alignedflowdata}Periodically developed friction factor data}

\hspace{-20mm}
\includegraphics[scale = 1.00]{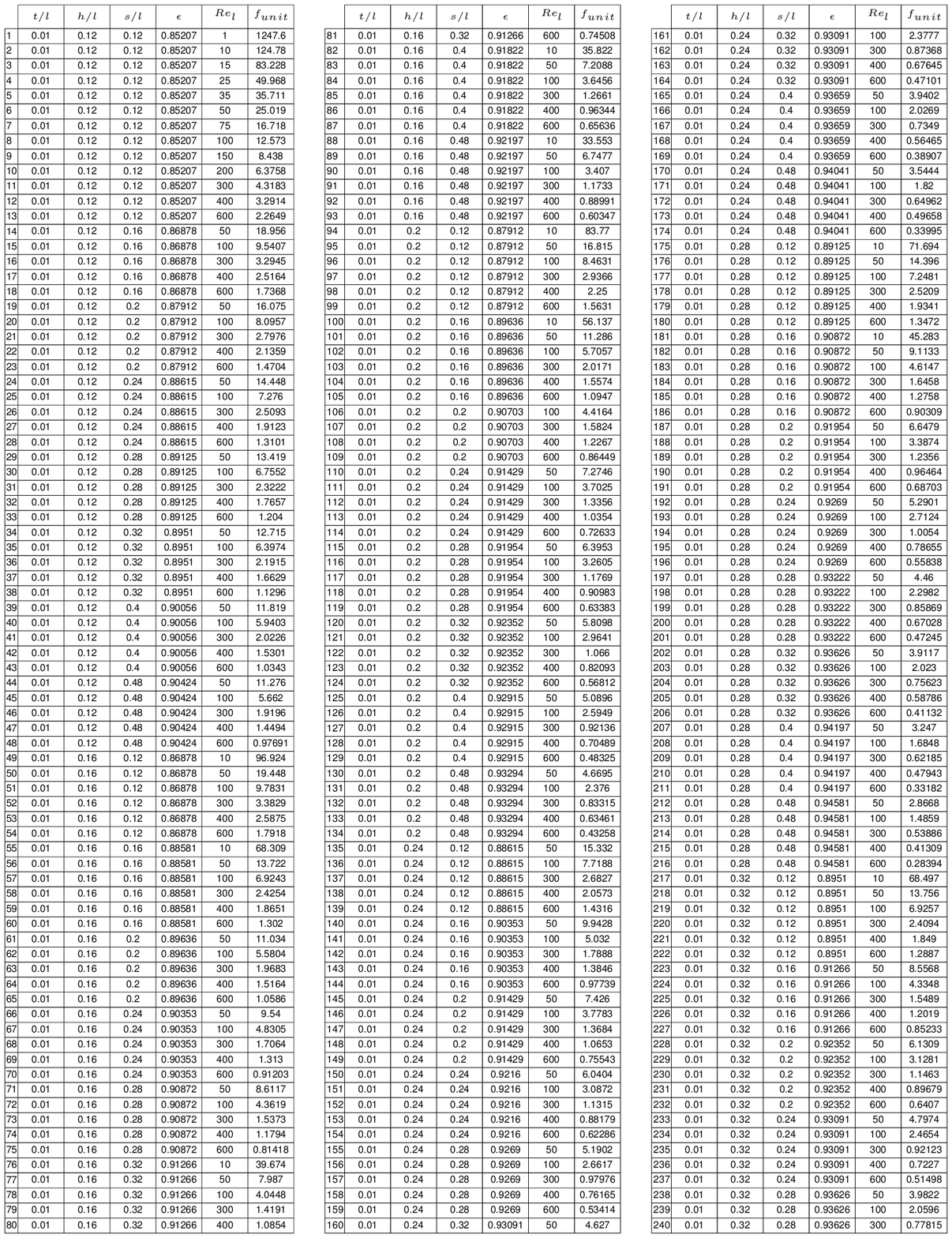}
\newpage
\clearpage
\hspace{-20mm}
\includegraphics[scale = 1.00]{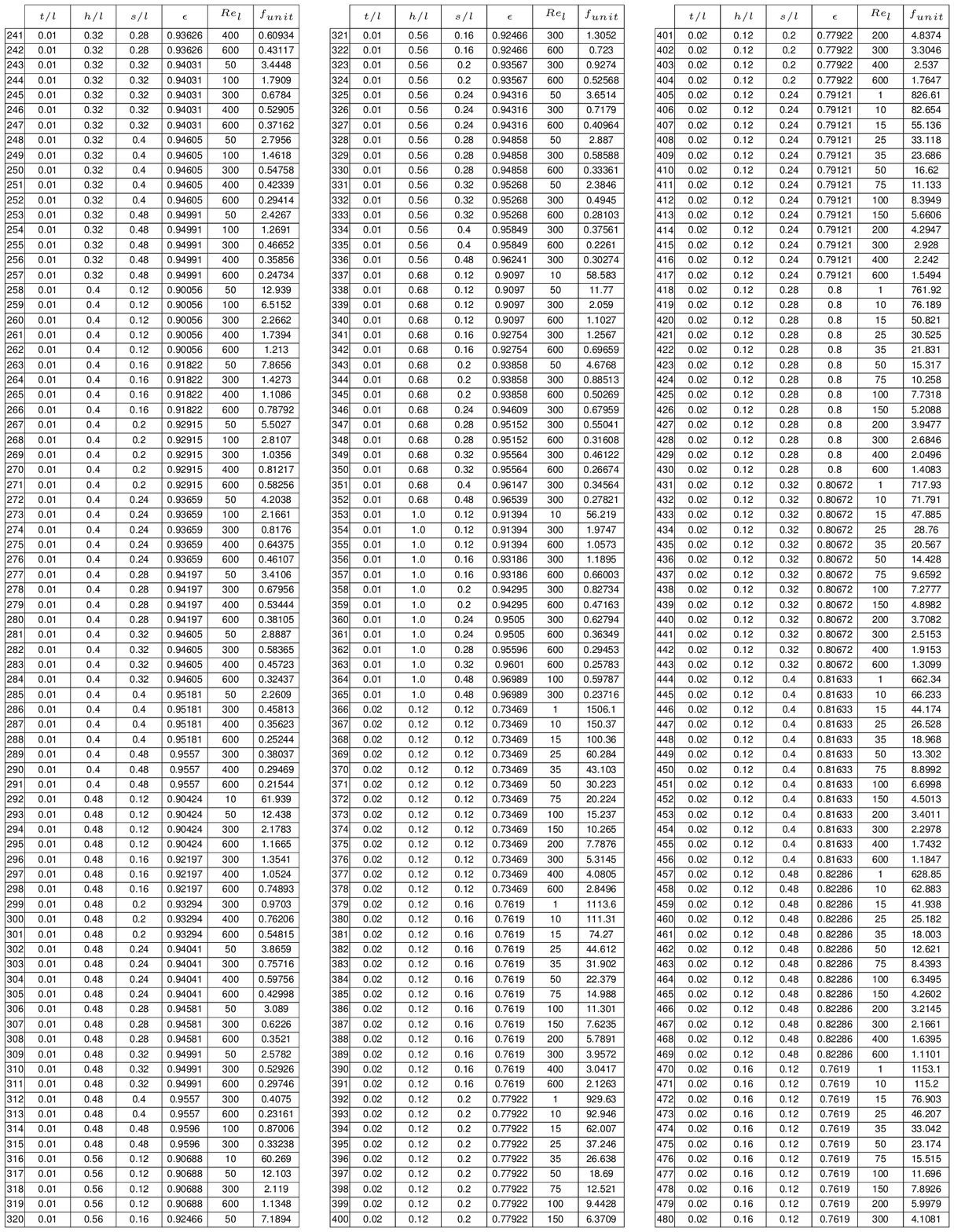}
\newpage
\clearpage
\hspace{-20mm}
\includegraphics[scale = 1.00]{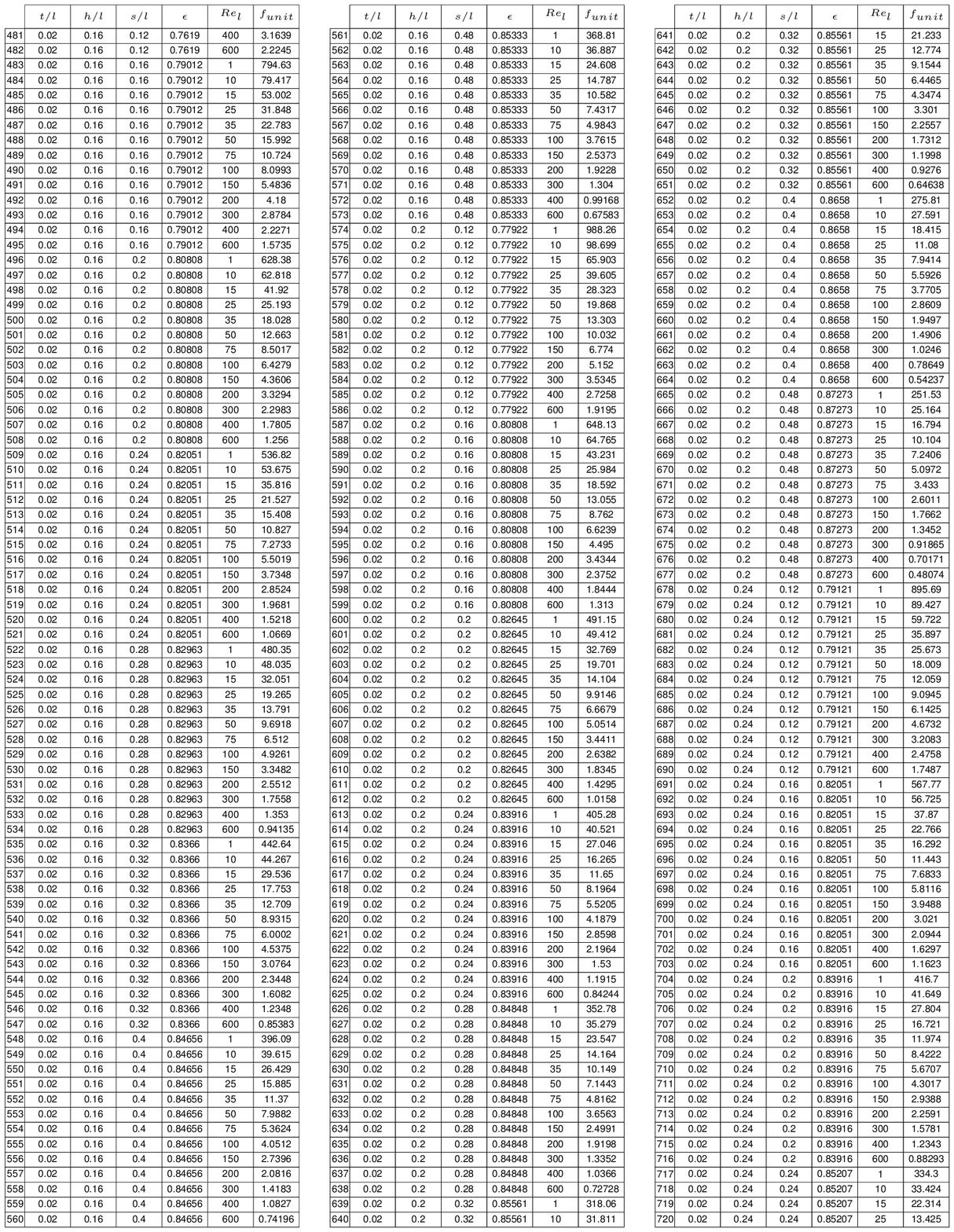}
\newpage
\clearpage
\hspace{-20mm}
\includegraphics[scale = 1.00]{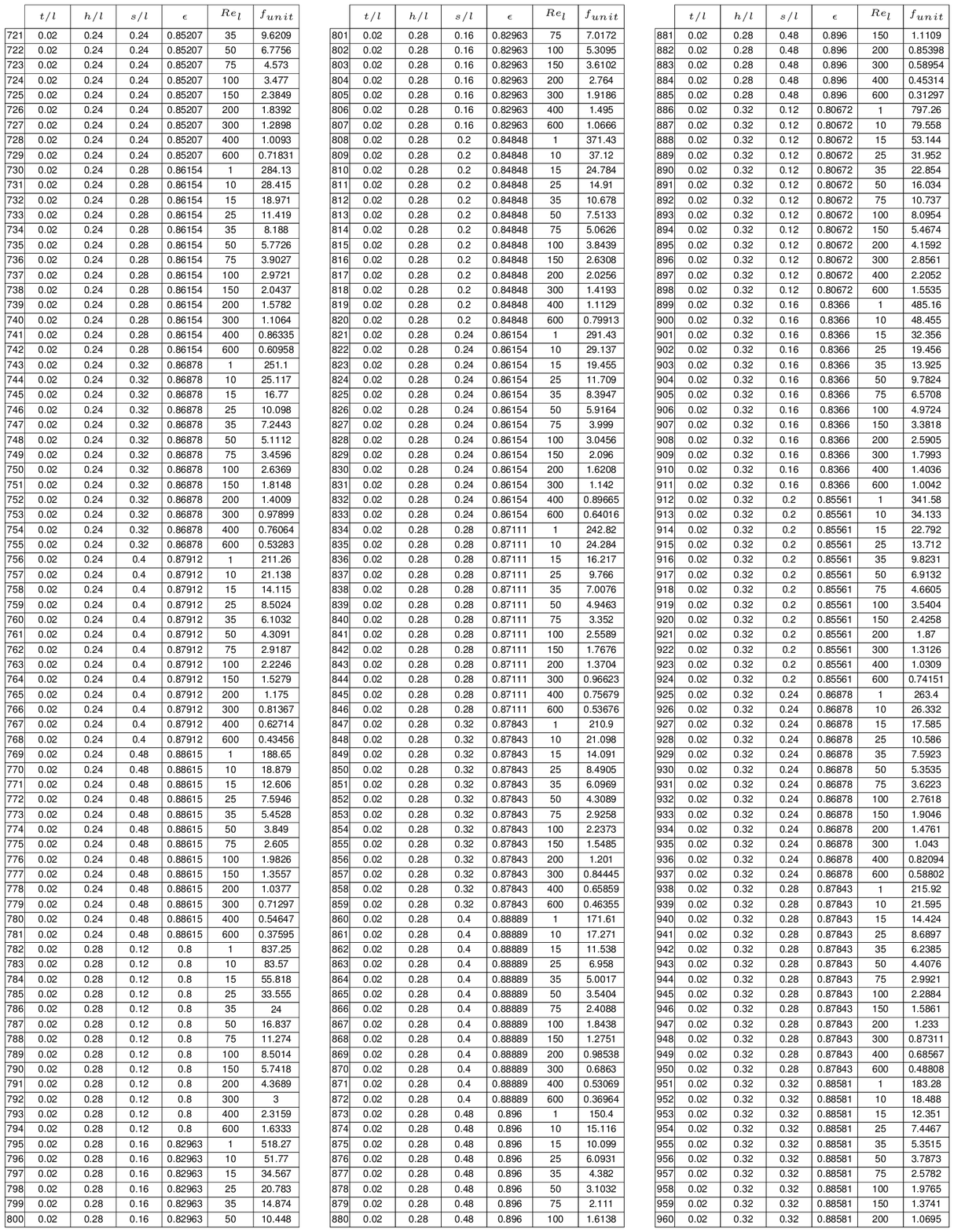}
\newpage
\clearpage
\hspace{-20mm}
\includegraphics[scale = 1.00]{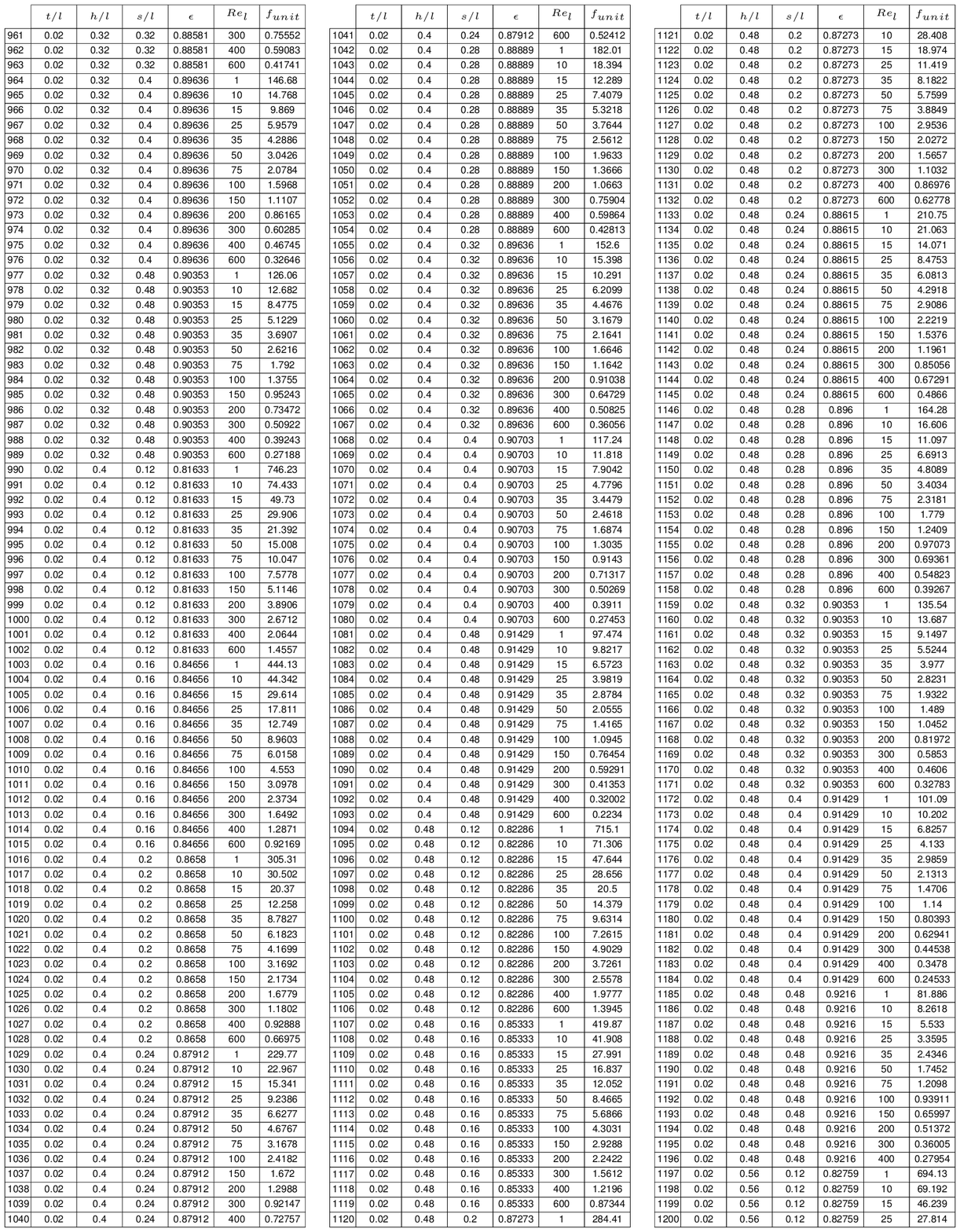}
\newpage
\clearpage
\hspace{-20mm}
\includegraphics[scale = 1.00]{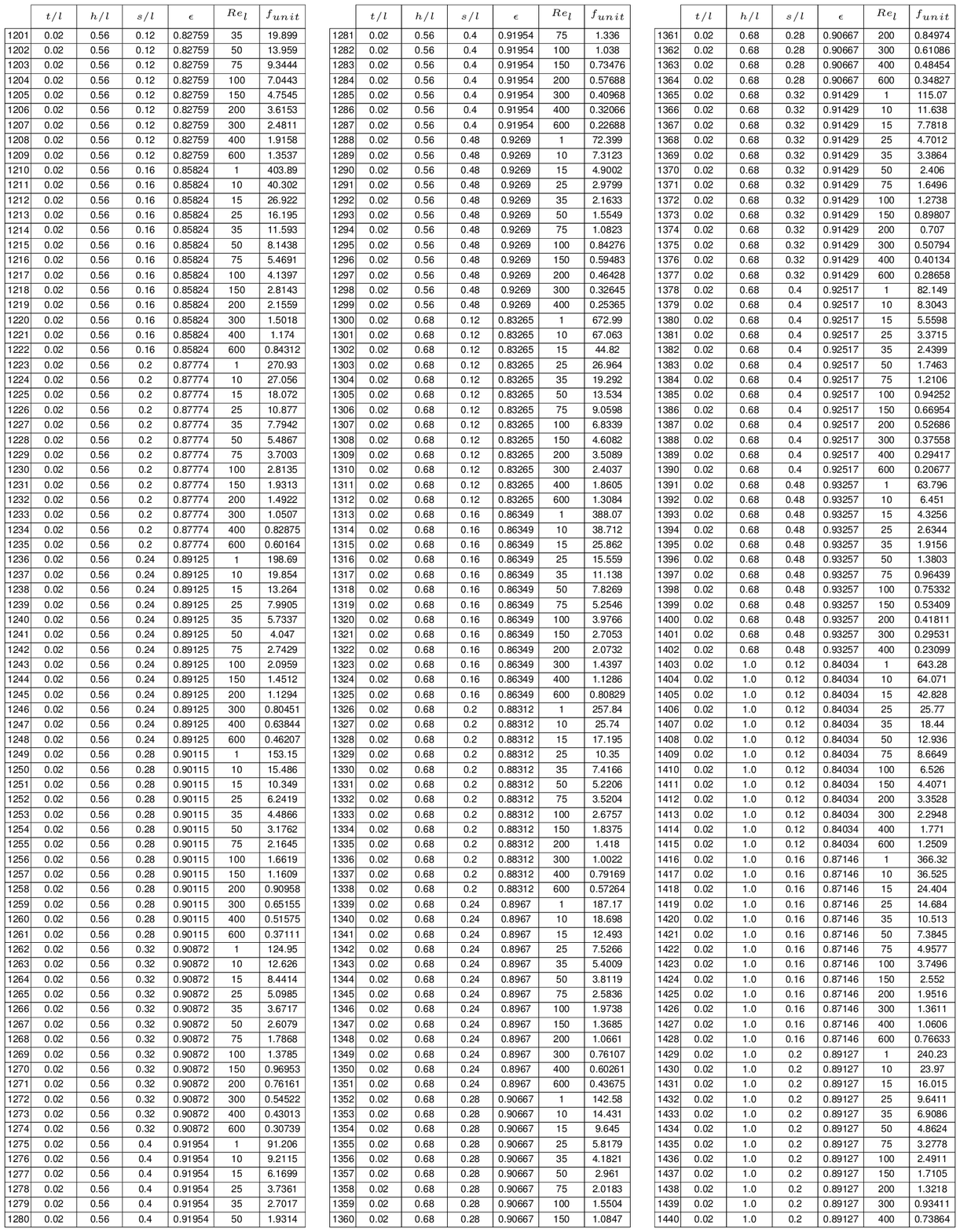}
\newpage
\clearpage
\hspace{-20mm}
\includegraphics[scale = 1.00]{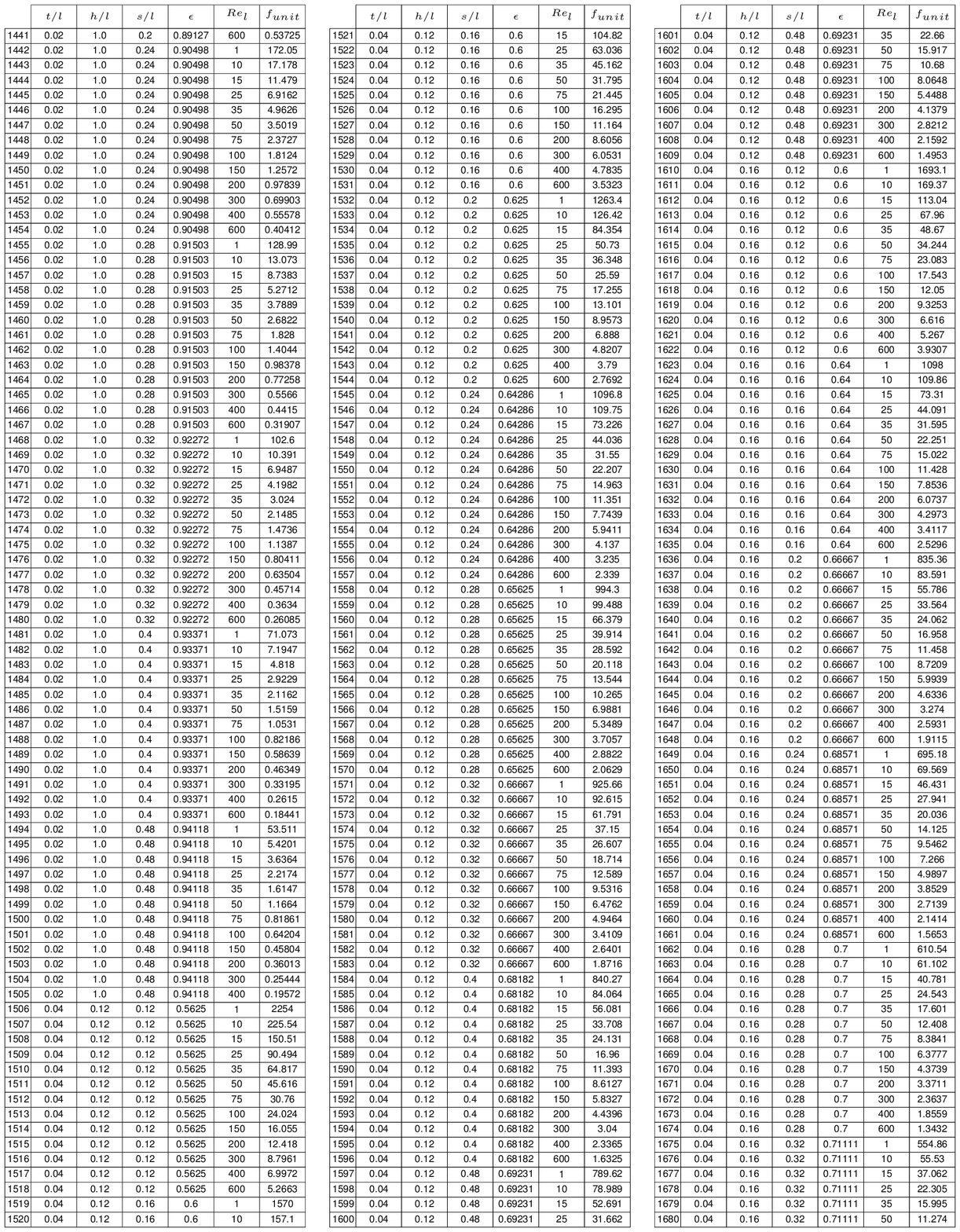}
\newpage
\clearpage
\hspace{-20mm}
\includegraphics[scale = 1.00]{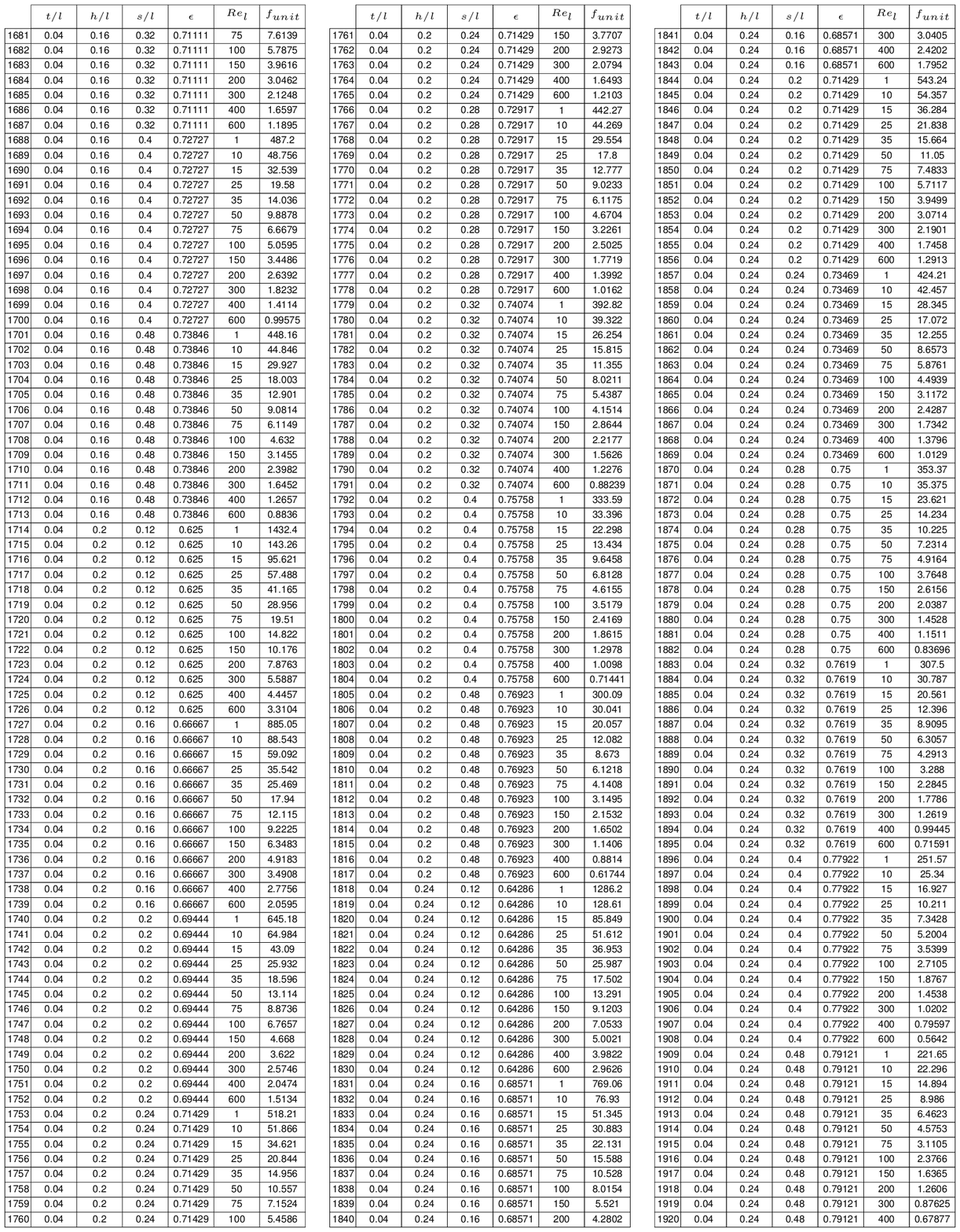}
\newpage
\clearpage
\hspace{-20mm}
\includegraphics[scale = 1.00]{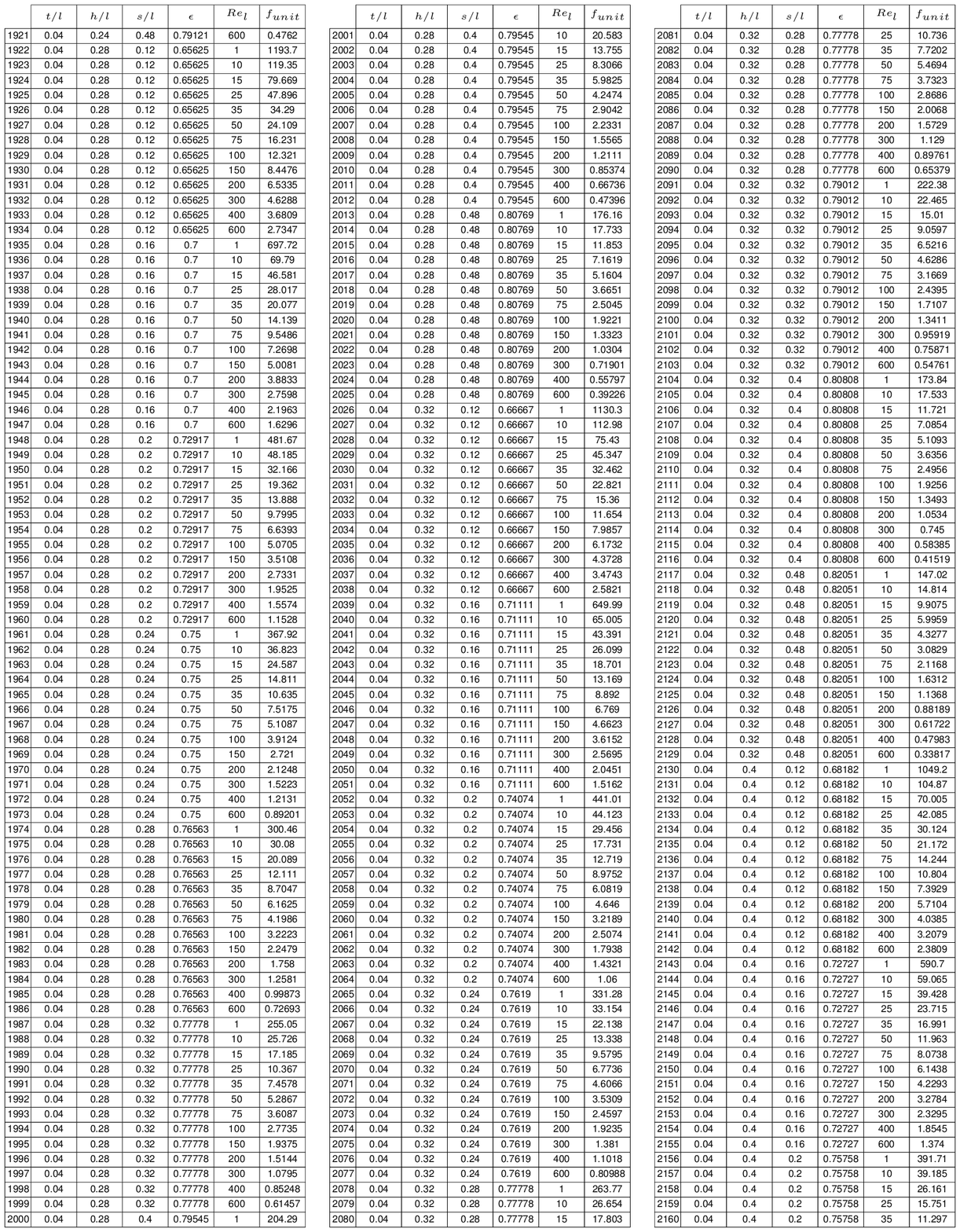}
\newpage
\clearpage
\hspace{-20mm}
\includegraphics[scale = 1.00]{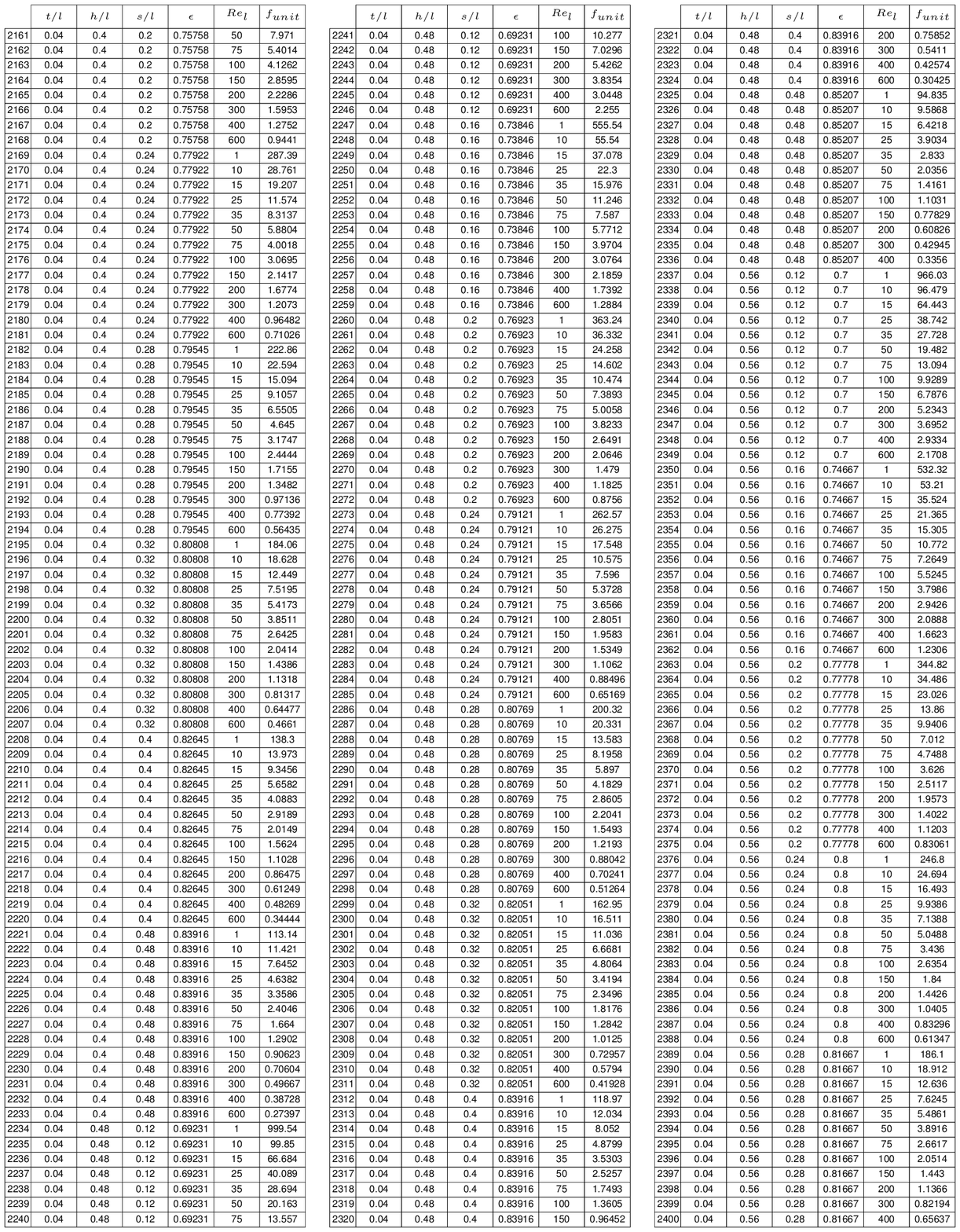}
\newpage
\clearpage
\hspace{-20mm}
\includegraphics[scale = 1.00]{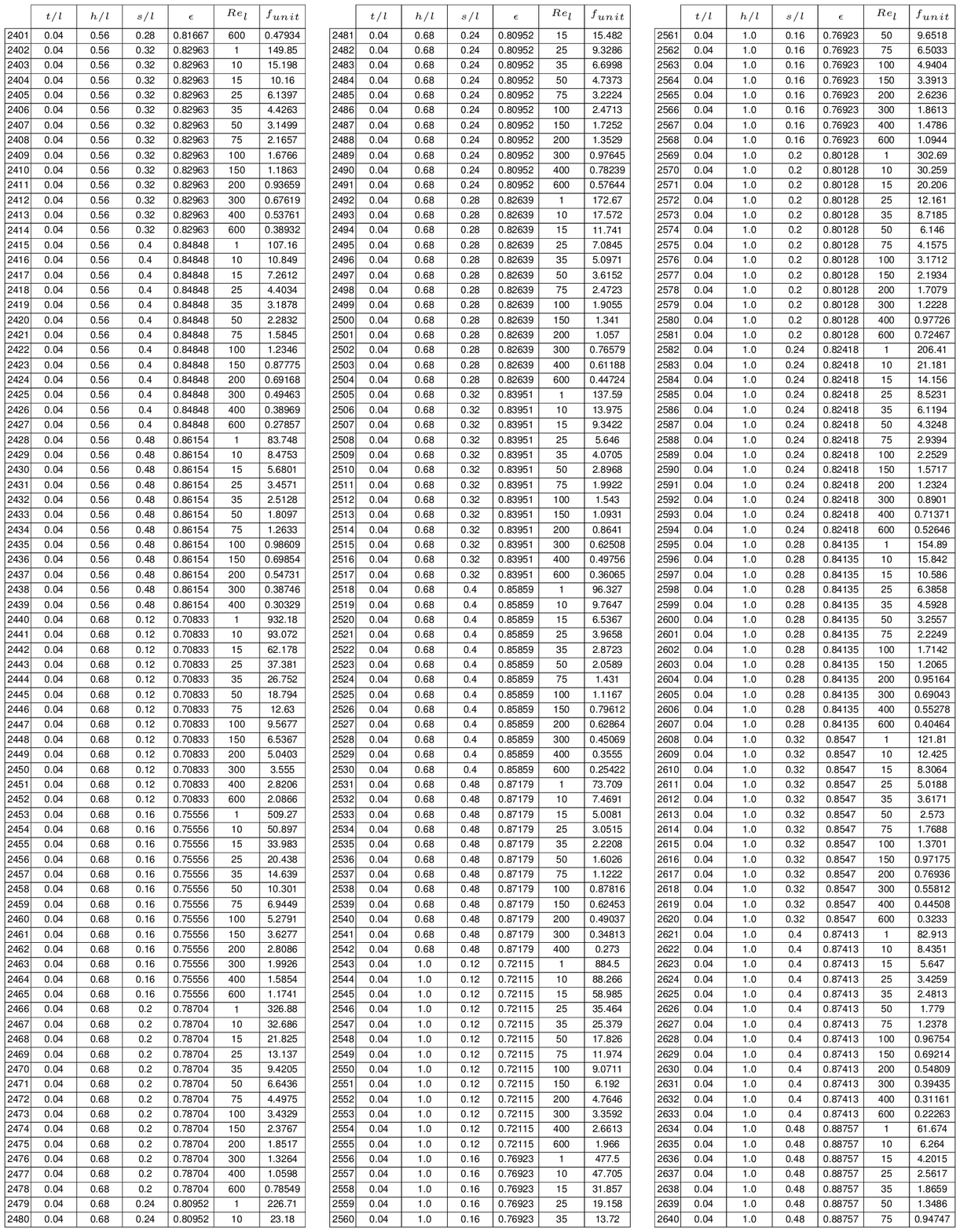}
\newpage
\clearpage
\hspace{-20mm}
\includegraphics[scale = 1.00]{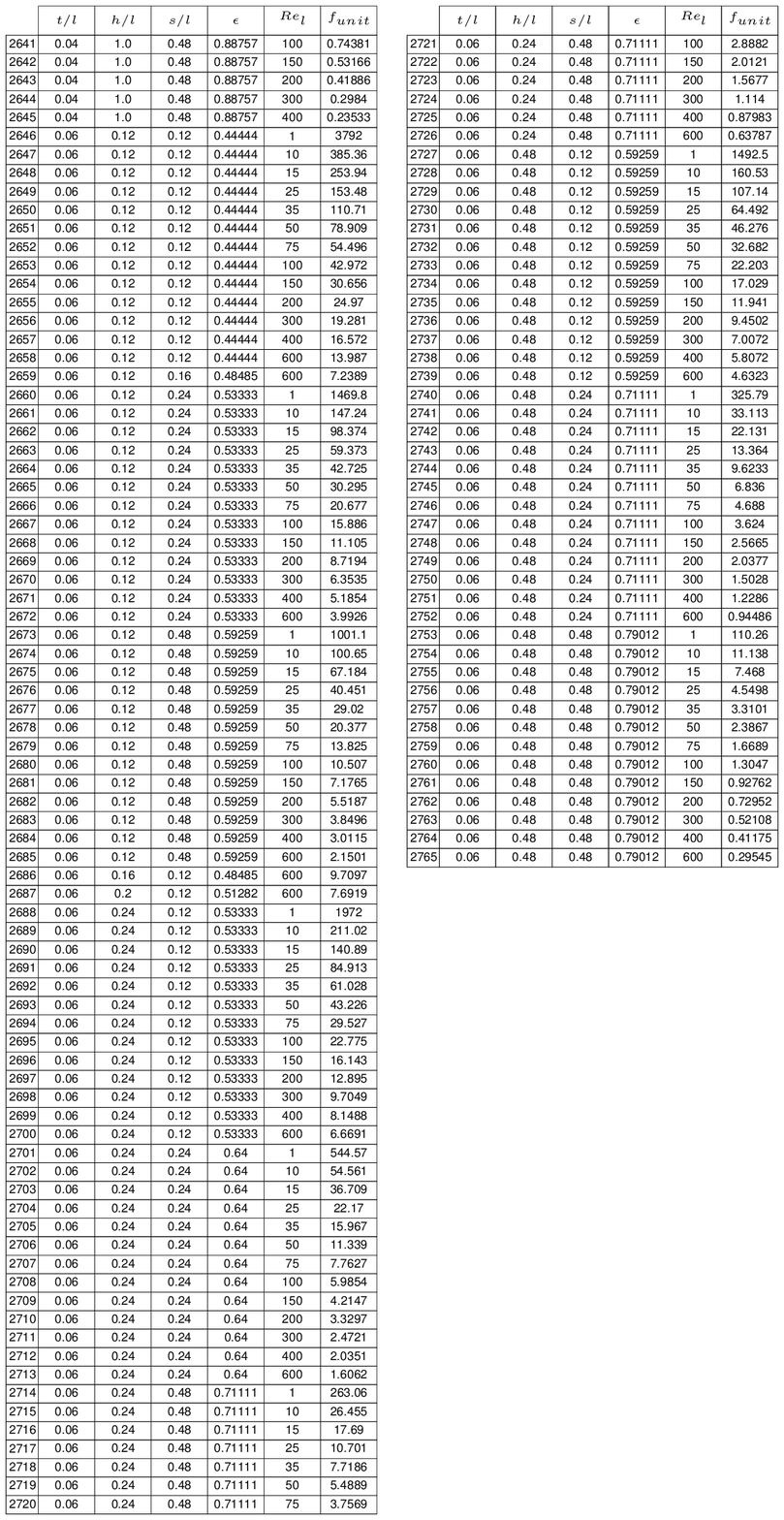}

\newpage
\clearpage

\nocite{*}
\bibliography{aipsamp.bib}% Produces the bibliography via BibTeX.

\end{document}